\documentclass[fleqn,usenatbib]{mnras}

\usepackage{amsmath}
\usepackage{amssymb}
\usepackage{graphicx}
\usepackage{mathptmx}
\usepackage{txfonts}
\usepackage[T1]{fontenc}

\newcommand{\ain}{a_\mathrm{in}}

\newcommand{\camb}{{\textsc{camb}}}

\newcommand{\CLpp}{C_L^{\phi\phi}}
\newcommand{\ClTT}{C_\ell^{\Theta\Theta}}
\newcommand{\delD}{\delta^\mathrm{(D)}}

\newcommand{\EQ}[1]{Eq.~(\ref{#1})}
\newcommand{\EQS}[2]{Eqs.~(\ref{#1}-\ref{#2})}
\newcommand{\fbtil}{{\tilde f}_\mathrm{b}}
\newcommand{\fbtilhat}{{\hat {\tilde f}_\mathrm{b}}}
\newcommand{\fcb}{f_\mathrm{cb}}

\newcommand{\flamingo}{{\tt FLAMINGO}}
\newcommand{\fnu}{f_\nu}
\newcommand{\halofit}{{\tt Halofit}}
\newcommand{\Hc}{{\mathcal H}}
\newcommand{\Hco}{{\mathcal H}_0}
\newcommand{\healpix}{{\textsc{HEALPix}}}
\newcommand{\hyphi}{{\tt hyphi}}

\newcommand{\Ob}{\Omega_\mathrm{b}}

\newcommand{\Obo}{\Omega_{\mathrm{b},0}}

\newcommand{\Om}{\Omega_\mathrm{m}}

\newcommand{\Omo}{\Omega_{\mathrm{m},0}}

\newcommand{\Ono}{\Omega_{\nu,0}}
\newcommand{\onu}{\omega_\nu}
\newcommand{\Pcb}{P_{\rm cb}}

\newcommand{\Pm}{P_{\rm m}}
\newcommand{\Pmdmo}{P_{\rm m,dmo}}
\newcommand{\Pmhyd}{P_{\rm m,hyd}}
\newcommand{\polspice}{{\textsc{polspice}}}
\newcommand{\SPk}{{\tt SP(k)}}
\newcommand{\swift}{{\textsc{swift}}}

\newcommand{\FDdNSl}{L1\_m10\_DMO}
\newcommand{\FDdNS}{L1\_m9\_DMO}
\newcommand{\FDdNSh}{L1\_m8\_DMO}
\newcommand{\FDdNSb}{L2p8\_m9\_DMO}
\newcommand{\FDdNSbb}{L5p6\_m10\_DMO}
\newcommand{\FDdNSbbb}{L11p2\_m11\_DMO}
\newcommand{\FDpNS}{Planck\_DMO}
\newcommand{\FDpNM}{PlanckNu0p12Var\_DMO}
\newcommand{\FDpNL}{PlanckNu0p24Fix\_DMO}
\newcommand{\FHdNS}{L1\_m9}
\newcommand{\FHdNSh}{L1\_m8}
\newcommand{\FHdNSb}{L2p8\_m9}
\newcommand{\FHpNS}{Planck}
\newcommand{\FHpNL}{PlanckNu0p24Fix}
\newcommand{\FHwdNS}{fgas+$2\sigma$}
\newcommand{\FHsdNS}{fgas-$2\sigma$}
\newcommand{\FHssdNS}{fgas-$4\sigma$}
\newcommand{\FHsssdNS}{fgas-$8\sigma$}
\newcommand{\FHjdNS}{Jet}
\newcommand{\FHjsdNS}{Jet\_fgas-$4\sigma$}
\newcommand{\FHSsdNS}{M$*-\sigma$}
\newcommand{\FHSsAsdNS}{M$*-\sigma$\_fgas-$4\sigma$}

\title{Non-linear CMB lensing with neutrinos and baryons: \flamingo{} simulations vs.~fast approximations}
\author[Amol Upadhye, et al.]{%
Amol Upadhye,$^1$ 
Juliana Kwan,$^1$
Ian G.~McCarthy,$^1$ 
Jaime Salcido,$^1$
John C.~Helly,$^2$
Roi Kugel,$^3$
\newauthor
Matthieu Schaller,$^{4,3}$
Joop Schaye,$^3$
Joey Braspenning,$^3$
Willem Elbers,$^2$
Carlos S.~Frenk,$^2$
\newauthor
Marcel P.~van Daalen,$^3$
Bert Vandenbroucke$^3$,
Jeger C.~Broxterman$^{4,3}$
\\
$^1$ Astrophysics Research Institute, Liverpool John Moores University, 146 Brownlow Hill, Liverpool L3 5RF, United Kingdom\\
$^2$ Institute for Computational Cosmology, Department of Physics, University of Durham, South Road, Durham, DH1 3LE, UK\\
$^3$ Leiden Observatory, Leiden University, PO Box 9513, 2300 RA Leiden, The Netherlands\\
$^4$Lorentz Institute for Theoretical Physics, Leiden University, PO Box 9506, NL-2300 RA Leiden, The Netherlands
}

\begin{document}
\label{firstpage}
\pagerange{\pageref{firstpage}--\pageref{lastpage}}
\maketitle

\begin{abstract}
Weak lensing of the cosmic microwave background is rapidly emerging as a powerful probe of neutrinos, dark energy, and new physics.  We present a fast computation of the non-linear CMB lensing power spectrum which combines non-linear perturbation theory at early times with power spectrum emulation using cosmological simulations at late times.  Comparing our calculation with lightcones from the \flamingo{} $5.6$~Gpc cube dark-matter-only simulation, we confirm its accuracy to $1\%$ ($2\%$) up to multipoles $L=3000$ ($L=5000$) for a $\nu\Lambda$CDM cosmology consistent with current data.  Clustering suppression due to small-scale baryonic phenomena such as feedback from active galactic nuclei can reduce the lensing power by $\sim 10\%$.  To our perturbation theory and emulator-based calculation we add \SPk, a new  fitting function for this suppression, and confirm its accuracy compared to the \flamingo{} hydrodynamic simulations to $4\%$ at $L=5000$, with similar accuracy for massive neutrino models.  We further demonstrate that scale-dependent suppression due to neutrinos and baryons approximately factorize, implying that a careful treatment of baryonic feedback can limit biasing neutrino mass constraints.
\end{abstract}

\section{Introduction}
\label{sec:introduction}

Cosmological bounds on the sum of neutrino masses, $M_\nu = \sum m_\nu \leq 0.12$~eV, are rapidly converging on the lower bound $M_\nu=0.06$~eV from laboratory experiments, promising a measurement in the next several years under restrictive assumptions about the cosmological model~\citep{deSalas:2017kay,Capozzi:2018ubv,Esteban:2020cvm,Planck:2018vyg,Palanque-Delabrouille:2019iyz,eBOSS:2020yzd,DES:2021wwk}.   However, relaxing these assumptions by allowing greater variation in the neutrino sector, dark energy, or scale-dependent bias, or by considering different combinations of probes, can substantially weaken cosmological bounds to $M_\nu \lesssim 0.5$~eV~\citep{Upadhye:2017hdl,DiValentino:2019dzu,DiValentino:2021imh,Sgier:2021bzf}, while relatively cosmology-independent terrestrial experiments are consistent with up to $M_\nu \sim 2$~eV~\citep{KATRIN:2019yun,KATRIN:2021uub}.  Meanwhile, tensions in the Hubble expansion rate and small-scale clustering demand a deeper understanding of the interdependent constraints on neutrinos and other components of the standard cosmological model~\citep{Leauthaud:2016jdb,Cai:2022dkh,Amon:2022ycy,Abdalla:2022yfr}.

Weak gravitational lensing of the cosmic microwave background (CMB), or CMB lensing, is a promising approach to measuring $M_\nu$.  It is immune to scale-dependent galaxy biasing, which has the potential to bias $M_\nu$ measurements from galaxy redshift surveys.  Its source, the CMB, is well-understood, consisting of photons which last scattered over a very narrow range of redshifts around $1100$.  Additionally, unlensed CMB temperature and polarization perturbations are very accurately characterized using linear perturbation theory applied to adiabatic, nearly-scale-invariant Gaussian random density fluctuations.  Thus, CMB lensing evades biases due to intrinsic alignments and photometric redshift errors that affect lower-redshift weak lensing surveys \citep{Weinberg:2013agg}. 

Next-generation CMB surveys are expected to improve upon the signal-to-noise ratio of the Planck survey by over an order of magnitude \citep{CMB-S4:2016ple,SimonsObservatory:2018koc,Liu:2022beb}.  Aside from mapping the CMB temperature on smaller scales, they will substantially advance our knowledge of the polarized CMB.  Systematic uncertainties due to astrophysical foregrounds and atmospheric noise are significantly smaller for the polarization than the temperature, and the lack of a small-scale primordial B mode polarization reduces the uncertainty due to cosmic variance.  Thus the next generation of experiments will be able to quantify CMB lensing to Legendre moments $L$ of a few thousand, or angles below $10$~arcmin.

Precision prediction of CMB lensing at these scales requires an understanding of non-linear corrections to the clustering of matter.  Several approaches to non-linear clustering have been explored in recent years.  Non-linear perturbation theory typically begins with the continuity and Euler equations of fluid dynamics, whose non-linear terms couple different Fourier modes together; see \citet{Crocce:2005xy,McDonald:2006hf,Taruya:2007xy,Matsubara:2007wj,Matsubara:2008wx,Pietroni:2008jx,Lesgourgues:2009am}.  We focus here on the Time-Renormalization Group (Time-RG) perturbation theory of \citet{Pietroni:2008jx,Lesgourgues:2009am}, designed for massive-neutrino cosmologies with scale-dependent clustering growth, as implemented in the {\tt redTime} code of \citet{Upadhye:2017hdl}.  Another approach to non-linear corrections begins with the halo model of clustering, detailed in \citet{Ma:2000ik,Seljak:2000gq,Cooray:2002dia}, tuned or supplemented by fitting functions to agree with large computer simulations, as in \citet{Smith:2002dz,Bird:2011rb}; \mbox{\citet{Takahashi:2012em};} \citet{Mead:2015yca,Mead:2020vgs}.  As the \halofit{} function of \citet{Bird:2011rb} was fit to neutrino simulations, we also consider it here.  Finally, the most accurate estimates of non-linear clustering, within limited ranges of parameters and redshifts, are emulators based upon large suites of N-body simulations; see \citet{Heitmann:2009cu,Lawrence:2009uk,Euclid:2020rfv,Moran:2022iwe}.  In the present study, we use Euclid Emulator 2 of \citet{Euclid:2020rfv} and the Mira-Titan IV emulator of \citet{Moran:2022iwe}.

As we are particularly interested in the small-scale ($k \gtrsim 0.1~h$~Mpc) suppression of clustering, hence CMB lensing, due to massive neutrinos, we must distinguish this suppression from the small-scale hydrodynamic effects of baryons.  Cooling and clumping of baryons leads to the formation of supernovae, which expel baryonic matter from galaxies.  Baryonic clustering at the centers of large halos feeds Active Galactic Nuclei (AGN), which, in turn, heat the baryonic gas and expel baryonic particles.  The combined effect of these phenomena is a suppression of clustering on $\sim 1$~Mpc scales, while  hydrodynamic models typically predict enhanced clustering on much smaller scales.  We model hydrodynamic effects through an innovative fitting function, \SPk{}, by \citet{Salcido:2023etz}, which uses the fact that the total hydrodynamic suppression is strongly correlated with the baryonic content of halos of a characteristic mass;\footnote{\SPk{} is publicly available at \href{https://github.com/jemme07/pyspk}{\tt github.com/jemme07/pyspk}~.} see also~\citet{vanDaalen:2019pst,Pandey:2023wqp}.  We briefly explore a one-parameter generalization of \SPk{}, showing that it covers a wide range of hydrodynamic models.

Our goal in this work is a fast, accurate computation of CMB lensing in the non-linear regime, quantified by the power spectrum of the lensing potential $\phi$, whose gradient determines the angle by which a CMB photon is deflected.  We combine the Mira-Titan IV emulator at low redshifts with Time-RG perturbation theory at high redshifts to yield a rapid computation of the matter power spectrum that is accurate at the times and length scales necessary for quantifying CMB lensing.  Baryonic feedback effects are modelled using \SPk{}.  Running in under a second, our calculation converges to sub-percent-level precision for Legendre moments $L \leq 10000$.  We release our code, \hyphi{}, publicly at \href{https://github.com/upadhye/hyphi}{\tt github.com/upadhye/hyphi}~.

Rigorously quantifying the errors in \hyphi{} requires a set of high-resolution, large-volume numerical simulations.  The \flamingo{} simulation suite of \citet{Schaye:2023jqv,Kugel:2023wte} is an ideal testing ground for \hyphi{} CMB lensing calculations.  With cosmological parameters chosen to match data from CMB probes and galaxy surveys, its CMB lensing power spectrum closely matches state-of-the-art measurements, as we show below.  \flamingo{} includes the largest-particle-number hydrodynamic simulation reaching $z=0$, which is necessary for covering the range of redshifts contributing the most to CMB lensing, and its relatively high mass resolution of $7\times 10^9~M_\odot$ provides a wealth of information on the impact of non-linear clustering.  Further, the \flamingo{} suite independently varies the neutrino masses and hydrodynamic feedback, both of which suppress small-scale clustering, allowing us to investigate their separate effects on lensing. 

We find close agreement between \hyphi{} and the \flamingo{} simulations across a wide range of neutrino masses, $0.06~{\rm eV} \leq M_\nu \leq 0.24$~eV; source redshift bins spanning $0 \leq z \leq 25$; and a variety of feedback models.  In particular, comparison with the \flamingo{} $M_\nu=0.06$~eV $5.6$~Gpc-box dark-matter-only (DMO) simulation demonstrates the accuracy of \hyphi{} to $1\%$ up to $L=3000$ and $2\%$ up to $L=5000$.  Dividing the range $0 \leq z \leq 25$ into eight source mass bins, we find $5\%$ agreement to at least $L=4000$ for each bin.  \flamingo{} simulations with larger $M_\nu$ and hydrodynamic feedback are run in smaller $1$~Gpc boxes, meaning that sample variance contributes more to the discrepancy between \hyphi{} and simulations, but even so, the two agree to $4\%$ up to $L=4000$ in a model with $M_\nu=0.24$~eV and the standard hydrodynamic feedback.  Furthermore, we demonstrate that neutrino and baryonic suppression of the lensing power factorize, facilitating a marginalization over feedback parameters to constrain $M_\nu$.  We show that this result extends to a wide variety of feedback models, including those with jets, as well as those reducing the cluster gas fraction well below its best-fit value.

This article is organized as follows.  Section~\ref{sec:background} provides overviews of CMB lensing, the \flamingo{} suite of simulations, and our modeling of baryonic feedback.  DMO models are considered in Sec.~\ref{sec:non-linear_cmb_lensing}, which studies the dependence of CMB lensing on the matter clustering, quantifies its suppression by massive neutrinos, and compares our fast computation to \flamingo{}.  Baryonic effects are included in Sec.~\ref{sec:hydrodynamic_suppression}, which demonstrates that these two types of scale-dependent suppression factorize.  Finally, Sec.~\ref{sec:conclusion} shows our conclusions.

\section{Background}
\label{sec:background}

\subsection{CMB lensing}
\label{subsec:bkg:cmb_lensing}

For a thorough, modern review of CMB lensing, see \citet{Lewis:2006fu}, which we briefly outline here.  Lensing may be understood as deflecting a light ray from the surface of last scattering as it passes through density perturbations along our past light cone.  The result is that a ray incident on a detector at angle $\hat n$ initially came from an angular position $\hat n + \vec \epsilon$ on the sky.  CMB temperature perturbations $\Theta(\hat n) := \delta T(\hat n) / \bar T$ on the last-scattering surface therefore are distorted by lensing to the observed perturbations $\tilde\Theta(\hat n) = \Theta(\hat n + \vec \epsilon)$.

Qualitatively, the effects of lensing on the CMB temperature and polarization power spectra split into three regimes, defined by large, intermediate, and small scales.  On large scales, with Legendre moments $\ell \ll 300$, lensing has only a small effect on perturbations.  Acoustic peaks on intermediate scales $\ell \sim 1000$ are smeared out by lensing.  On small scales, $\ell \gg 3000$, where diffusion damping sharply reduces the power of the unlensed CMB perturbations, lensing and other secondary anisotropies dominate the power.

At each position along a photon's path to us, its deflection is proportional to the gradient of the local gravitational potential.  Thus its total deflection is the gradient of the line-of-sight integral of the gravitational potential, weighted by a lensing kernel $g$ defined below.  This integral, known as the lensing potential $\phi(\hat n)$, sources the deflection $\vec \epsilon = \vec\nabla \phi$.

Gravitational lensing of the CMB may be quantified statistically using the power spectrum $\CLpp$ of $\phi$.  In the Limber approximation of \citet{Limber:1953}, appropriate to the scales of interest to us here, this is
\begin{eqnarray}
  \CLpp
  &\approx&\
  \int_0^{\chi_\star} d\chi \,
  P_\Phi\!\left(\frac{L}{\chi}, z(\chi)\right)
  \frac{g(\chi)^2}{\chi^6}
  \label{e:CLpp_los}
  \\
  P_\Phi(k,z)
  &=&
  \left[\frac{3\Om(z) \Hc(z)^2}{2k^2}\right]^2 \Pm(k,z)
  =
  \frac{9\Omo^2 \Hco^4}{4a^2 k^4} \Pm(k,z)~
  \label{e:P_Phi}
  \\
  g(\chi)
  &=&
  2\chi (1 - \chi/\chi_\star).
  \label{e:g_kernel}
\end{eqnarray}
Here, $P_\Phi$ and $\Pm$ are, respectively, the power spectra of the gravitational potential and total matter in units of volume; $\chi(z)$ is the comoving distance to redshift $z$; $\chi_\star = \chi(z_\star)$; and $z_\star$ is the redshift of the baryon drag epoch.  The cosmological parameters $\Omo$ and $\Hco$ are the present-day values of the matter density as a fraction of the critical density $\bar \rho_{\rm m,0} / \bar \rho_{\rm crit,0}$, and the derivative of the logarithm of the scale factor with respect to conformal time $d \log(a) / d \tau$, respectively.   The functions $\Hc(z) = d \log(a) / d \tau$ and $\Om(z) = \Omo \Hco^2 (1+z) / \Hc^2$ are the expansion rate and matter fraction at all redshifts.

In turn, the power spectrum of the lensed temperature perturbation $\tilde\Theta(\hat n)$ is found from the appropriately-weighted convolution of $\CLpp$ with the unlensed power $\ClTT$:
\begin{eqnarray}
  C_\ell^{\tilde\Theta \tilde\Theta}
  \!\!&=&
  \!\!\!\int \!\!\!\frac{d^2\vec L}{(2\pi)^2} 
  [\vec L \cdot(\vec\ell-\vec L)]^2
  C_{\ell}^{\Theta \Theta} C_{L}^{\phi\phi} 
  + \left(\!1-\ell^2\!\!\!\int \!dL \, L^3 C_L^{\phi\phi}\!\right)
  C_\ell^{\Theta \Theta}.\quad
\end{eqnarray}
Here we work in the flat-sky limit in which $\vec \ell$ is the two-dimensional Fourier conjugate to the direction $\hat n$.  Factors of $L$ and $\ell-L$ arise from gradients of the lensing potential power spectrum.  The first term on the right hand side describes the lensing-induced smearing of $\ClTT$, while the second represents a smooth suppression of power due to lensing.  Lensed CMB polarization power spectra are similarly found by convolution with $\CLpp$; see~\citet{Lewis:2006fu}.

Thus $\CLpp$ is necessary for quantifying lensing of the CMB, and is useful for comparing the observed lensing to theoretical predictions.  Our goal henceforth is a computation of $\CLpp$, including non-linear clustering as well as power suppression by massive neutrinos and hydrodynamic effects.

\subsection{The \flamingo{} simulations}
\label{subsec:bkg:flamingo_simulations}

The \flamingo{} simulations used here are thoroughly described in \citet{Schaye:2023jqv,Kugel:2023wte}.  They employed the \swift{} code of \citet{Schaller:2023hzn}, which includes gravitation, hydrodynamic feedback, and subgrid models for unresolved physics relevant to galaxy formation, including metal-dependent radiative cooling, star formation, stellar evolution, and stellar and AGN feedback.  Neutrinos were incorporated into the simulations through the $\delta f$ method of~\citet{Elbers:2020lbn}, and included in the initial conditions of {\textsc{MonofonIC}}, \citet{Hahn:2020lvr,2020ascl.soft08024H}, as detailed and implemented in \citet{Elbers:2022tvb,Elbers:2022xid,2022ascl.soft12004E}.

\begin{table*}
  \caption{
    List of \flamingo{} simulations, with simulation parameters and
    cosmological parameters.  $n_{\rm LC}$ is the number of high-redshift
    simulation lightcones produced; these cover the range
    $0 \leq z \leq 3$ for $L_{\rm sim} = 1000$~Mpc$/h$;
    $0 \leq z \leq 5$ for $L_{\rm sim} = 2800$~Mpc$/h$; and
    $0 \leq z \leq 25$ for $L_{\rm sim} = 5600$~Mpc$/h$.
    For feedback fits and standard deviations, as well as a description
    of the feedback models realizing the $f_{\rm gas}$ and SMF listed, see 
    \citet{Schaye:2023jqv,Kugel:2023wte}.
    \label{t:flamingo_sims}
  }
  \begin{tabular}{l|cccccccccccc}
    \hline
    Name
    & $L_{\rm sim}$[Mpc] & $N_{\rm CDM}^{1/3}$ & $N_\nu^{1/3}$ & $n_{\rm LC}$
    & $f_{\rm gas}$ & SMF 
    & $M_\nu$[eV] & $h$ & $\Obo$ & $\Omo$ & $10^9 A_s$ & $n_s$\\
    \hline
    \FDdNSl
    & $1000$ & $900$ & $500$ & $1$
    & N/A & N/A
    & $0.06$ & $0.681$ & $0.0486$ & $0.306$ & $2.099$ & $0.967$
    \\
    \FDdNS
    & $1000$ & $1800$ & $1000$ & $1$
    & N/A & N/A
    & $0.06$ & $0.681$ & $0.0486$ & $0.306$ & $2.099$ & $0.967$
    \\
    \FDdNSh
    & $1000$ & $3600$ & $2000$ & $1$
    & N/A & N/A
    & $0.06$ & $0.681$ & $0.0486$ & $0.306$ & $2.099$ & $0.967$
    \\
    \FDdNSb
    & $2800$ & $5040$ & $2800$ & $8$
    & N/A & N/A
    & $0.06$ & $0.681$ & $0.0486$ & $0.306$ & $2.099$ & $0.967$
    \\
    \FDdNSbb
    & $5600$ & $5040$ & $2800$& $8$
    & N/A & N/A
    & $0.06$ & $0.681$ & $0.0486$ & $0.306$ & $2.099$ & $0.967$
    \\
    \FDdNSbbb
    & $11200$ & $5040$ & $2800$& $8$
    & N/A & N/A
    & $0.06$ & $0.681$ & $0.0486$ & $0.306$ & $2.099$ & $0.967$
    \\
    \FDpNS
    & $1000$ & $1800$ & $1000$ & $1$
    & N/A & N/A
    & $0.06$ & $0.673$ & $0.0494$ & $0.316$ & $2.101$ & $0.966$
    \\
    \FDpNM
    & $1000$ & $1800$ & $1000$ & $1$
    & N/A & N/A
    & $0.12$ & $0.673$ & $0.0496$ & $0.316$ & $2.113$ & $0.967$
    \\
    \FDpNL
    & $1000$ & $1800$ & $1000$ & $1$
    & N/A & N/A
    & $0.24$ & $0.673$ & $0.0494$ & $0.316$ & $2.101$ & $0.966$
    \\
    \FHdNS
    & $1000$ & $1800$ & $1000$ & $1$
    & fit & fit 
    & $0.06$ & $0.681$ & $0.0486$ & $0.306$ & $2.099$ & $0.967$
    \\
    \FHdNSh
    & $1000$ & $3600$ & $2000$ & $1$
    & fit & fit 
    & $0.06$ & $0.681$ & $0.0486$ & $0.306$ & $2.099$ & $0.967$
    \\
    \FHdNSb
    & $2800$ & $5040$ & $2800$ & $8$
    & fit & fit 
    & $0.06$ & $0.681$ & $0.0486$ & $0.306$ & $2.099$ & $0.967$
    \\
    \FHpNS
    & $1000$ & $1800$ & $1000$ & $1$
    & fit & fit 
    & $0.06$ & $0.673$ & $0.0494$ & $0.316$ & $2.101$ & $0.966$
    \\
    \FHpNL
    & $1000$ & $1800$ & $1000$ & $1$
    & fit & fit 
    & $0.24$ & $0.673$ & $0.0494$ & $0.316$ & $2.101$ & $0.966$
    \\
    \FHwdNS
    & $1000$ & $1800$ & $1000$ & $1$
    & fit$+2\sigma$ & fit
    & $0.06$ & $0.681$ & $0.0486$ & $0.306$ & $2.099$ & $0.967$
    \\
    \FHsdNS
    & $1000$ & $1800$ & $1000$ & $1$
    & fit$-2\sigma$ & fit
    & $0.06$ & $0.681$ & $0.0486$ & $0.306$ & $2.099$ & $0.967$
    \\
    \FHssdNS
    & $1000$ & $1800$ & $1000$ & $1$
    & fit$-4\sigma$ & fit
    & $0.06$ & $0.681$ & $0.0486$ & $0.306$ & $2.099$ & $0.967$
    \\
    \FHsssdNS
    & $1000$ & $1800$ & $1000$ & $1$
    & fit$-8\sigma$ & fit
    & $0.06$ & $0.681$ & $0.0486$ & $0.306$ & $2.099$ & $0.967$
    \\
    \FHjdNS
    & $1000$ & $1800$ & $1000$ & $1$
    & fit${}^\dagger$ & fit${}^\dagger$
    & $0.06$ & $0.681$ & $0.0486$ & $0.306$ & $2.099$ & $0.967$
    \\
    \FHjsdNS
    & $1000$ & $1800$ & $1000$ & $1$
    & fit$-4\sigma^\dagger$ & fit${}^\dagger$
    & $0.06$ & $0.681$ & $0.0486$ & $0.306$ & $2.099$ & $0.967$
    \\
    \FHSsdNS
    & $1000$ & $1800$ & $1000$ & $1$
    & fit & fit$-1\sigma$
    & $0.06$ & $0.681$ & $0.0486$ & $0.306$ & $2.099$ & $0.967$
    \\
    \FHSsAsdNS
    & $1000$ & $1800$ & $1000$ & $1$
    & fit$-4\sigma$ & fit$-1\sigma$
    & $0.06$ & $0.681$ & $0.0486$ & $0.306$ & $2.099$ & $0.967$
    \\
    \hline
  \end{tabular}
  
    ${}^\dagger$ \FHjdNS{} and \FHjsdNS{} realize the fit $f_{\rm gas}$ and SMF
    listed using momentum jets rather than thermal energy injection, as
    detailed in \citet{Schaye:2023jqv}.
\end{table*}

Table~\ref{t:flamingo_sims} lists the code and cosmological parameters for the \flamingo{} simulations used in this work.  We will use the large-volume run \FDdNSbb{} to quantify the accuracy of our $\CLpp$ computation at high redshifts; the $L_{\rm sim}=2.8$~Gpc runs, \FDdNSb{} and \FHdNSb{}, to study the impact of baryonic feedback; and the several $L_{\rm sim}=1$~Gpc runs to asses the impact of different neutrino masses and feedback processes on $\CLpp$.

Lightcones are data products of the \flamingo{} simulations of \citet{Schaye:2023jqv}.  Lightcone maps apply the \healpix{} pixelization of \citet{Gorski:2004by} to spherical shells, recording the total mass contained in each resulting volume element.  Shell thicknesses in redshift space, $z_{i+1}-z_i$, are $0.05$ up to $z=3$, and then $0.25$ up to $z=5$, gradually increasing to $5$ at $z=15$, corresponding to homogeneous-universe comoving radii $\chi(z)$ between $\chi_i=\chi(z_i)$ and $\chi_{i+1}=\chi(z_{i+1})$. Each shell is divided into $12 N_\mathrm{side}^2 = 3,221,225,472$ pixels, each of angular size $166$~arcsec$^2$, with the \healpix{} parameter $N_\mathrm{side}=16,384$.  For each \healpix{} angular pixel $j$, the lensing convergence $\kappa_j$ is computed through a summation discretizing the line-of-sight integral over comoving distance:
\begin{equation}
  \kappa_j
  =
  \frac{3}{2} \Omo \Hco^2
  \sum_i \Delta\chi_i \,\, \left<\chi_i\right> (1+z_i)
  \left(1-\frac{\left<\chi\right>_i}{\chi_\star}\right)
  \frac{M_{ij} - {\bar M}_{ij}}{{\bar M}_{ij}}.
  \label{e:kappa}
\end{equation}
Here, $\Delta\chi_i := \chi_{i+1}-\chi_i$; $\left<\chi \right>_i$ is the mean comoving distance in redshift shell $i$; $\chi_\star$ is the comoving distance to the surface of last scattering; $M_{ij}$ is the mass in shell $i$ and \healpix{} pixel $j$; and ${\bar M}_{ij} = \frac{4\pi}{3} {\bar \rho}_{\mathrm{m},0} (\chi_{i+1}^3 - \chi_i^3) / (12 N_\mathrm{side}^2)$ is the homogenous-universe mass in $i$ and $j$.  We fix $\chi_\star = \chi(z_\star)$, with $z_\star = 1089.80$ as measured by \citet{Planck:2018vyg}.  

Standard power spectrum computation codes such as the \polspice{} code of \citet{Szapudi:2000xj} cannot process more than $2^{31}\approx 2\times 10^9$ pixels, so we downsample the $\kappa$ map to $N_\mathrm{side}=8192$.  Then we use \polspice{} with the {\tt -pixelfile YES} option to compute its power spectrum, which we subsequently multiply by $4/L^4$ to yield the lensing potential power spectrum $\CLpp$.  We confirm by comparison to power spectra from lower-resolution maps that our $\CLpp$ with $N_\mathrm{side}=8192$ has converged to better than $1\%$ ($2\%$) up to $L=5000$ ($L=6000$), which is sufficient for testing \hyphi{}.

\begin{figure}
  \begin{center}
    \includegraphics[width=90mm]{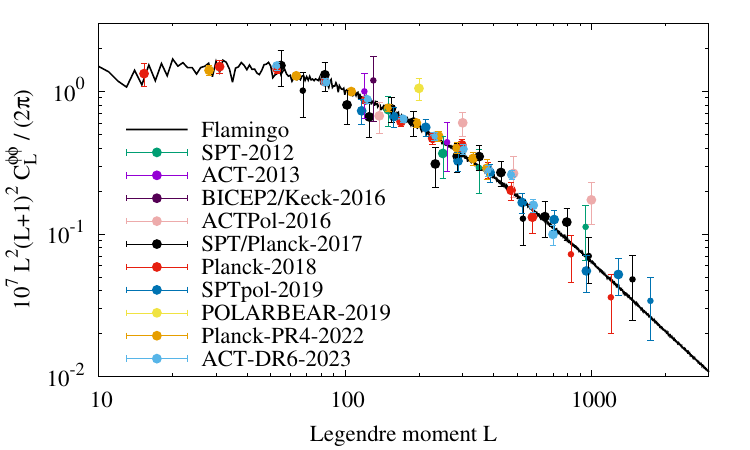}%
  \end{center}
  \caption{
    Lensing potential power spectrum from \flamingo{} model \FDdNSbb{} of
    Table~\ref{t:flamingo_sims}, integrated up to $z=25$ and averaged over eight
    lightcones, compared with measurements of
    \protect\citet{vanEngelen:2012va,Das:2013zf,BICEP2:2016rpt,Sherwin:2016tyf,Simard:2017xtw,Planck:2018lbu,Wu:2019hek,POLARBEAR:2019ywi,Carron:2022eyg,ACT:2023dou}.
    Small (large) data points show detections to $2 \sigma$ ($\geq 3 \sigma$).
    \label{f:CLpp_data_vs_flamingo}
  }
\end{figure}

The lensing potential power spectrum from the $L_{\rm sim}=5.6$~Gpc \flamingo{} simulation, \FDdNSbb{}, averaged over lightcones computed from eight different positions in the simulation volume, is compared with several recent observations in Fig.~\ref{f:CLpp_data_vs_flamingo}.  Agreement across two orders of magnitude in the Legendre moment is impressive, demonstrating that the \flamingo{} simulation suite is an appropriate tool for quantifying non-linear CMB lensing. Since we are primarily concerned with neutrino and baryonic suppression effects, which are largest at small scales $L \gtrsim 1000$ where data error bars are large, we henceforth focus on \flamingo{} directly as a means of quantifying these small-scale effects and testing their predictions. 

\begin{figure}
    \begin{center}
        \includegraphics[width=90mm]{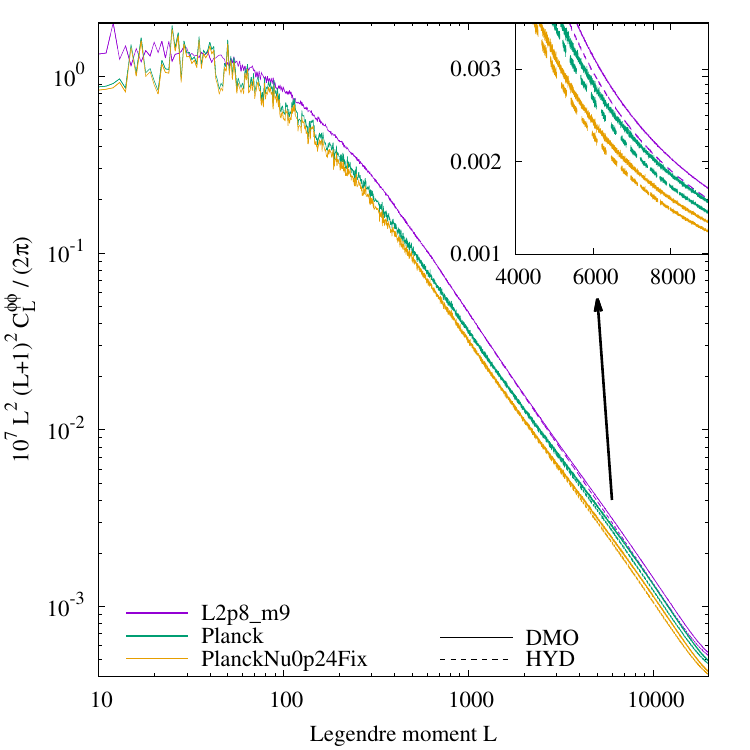}%
    \end{center}
    \caption{
      $\CLpp$ from \flamingo{} \FHdNSb{}, \FHpNS{}, and \FHpNL{} 
      simulations (dashed) and their corresponding DMO 
      simulations (solid) computed over the range of redshifts given in 
      Table~\ref{t:flamingo_sims}.
      (Inset) Baryonic suppression is evident at small scales relevant for
      next-generation probes.
      \label{f:CLpp_flamingo}
    }
\end{figure}

Figure~\ref{f:CLpp_flamingo} shows \flamingo{} lensing potential power spectra for several models.  Although neutrinos and baryons both suppress $\CLpp$ by $\sim 10\%$, these effects depend very differently upon scale. The neutrino suppression extends down to $L\approx 100$, while the baryonic suppression is visible only beyond $L=1000$. Quantifying the difference between these effects is a major goal of this article.

The other obvious difference between \FHdNSb{} and the Planck-like models, \FHpNS{} and \FHpNL{}, in Fig.~\ref{f:CLpp_flamingo}, is in their large-scale power.  This difference is due to the fact the the Planck-like models are simulated in smaller boxes whose lightcones are limited to $z\leq 3$, so our $\CLpp$ computation based upon \EQ{e:kappa} is limited to that redshift range.  A direct comparison of these power spectra to the data would first require some approximation of the higher-$z$ contribution to $\CLpp$, such as the use of the \flamingo{} matter power spectra in the line-of-sight integral of \EQS{e:CLpp_los}{e:g_kernel}.  However, since our goal is to test fast approximations using the lightcones, we take the opposite approach, and limit our perturbative and emulated computations to the redshift range covered by each lightcone.

\begin{figure}
    \begin{center}
        \includegraphics[width=90mm]{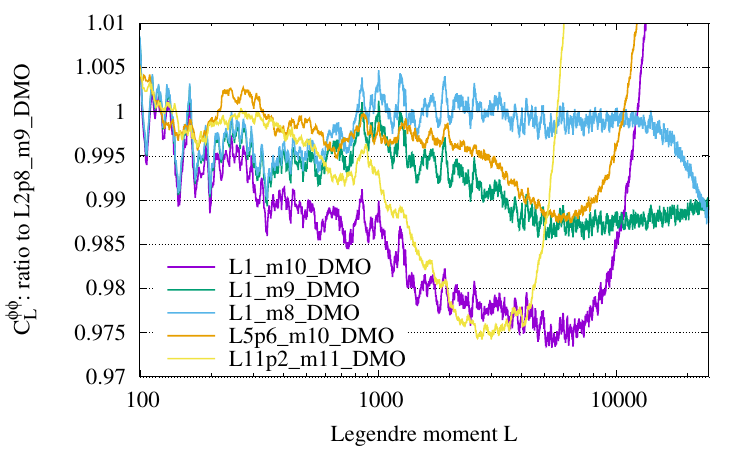}%
    \end{center}
    \caption{
    Ratios of $\CLpp$ computed from the \FDdNSl{}, \FDdNS{}, \FDdNSh{}, 
    \FDdNSbb{}, and \FDdNSbbb{} simulations to that from \FDdNSb{}, smoothed 
    using a centered $100$-point moving average. The higher-resolution runs
    \FDdNS{}, \FDdNSh{}, \FDdNSb{}, and \FDdNSbb{}, with multiple box sizes, 
    agree to $1\%$ up to $L=3000$ and $<1.5\%$ beyond $L=20000$.
    \label{f:CLpp_convergence}
    }
\end{figure}

One strength of the \flamingo{} simulation suite is that it contains multiple resolutions and box sizes for the same cosmology, allowing for convergence tests.  Figure~\ref{f:CLpp_convergence} compares $\CLpp$ integrated up to $z=3$, computed from six different \flamingo{} simulations, with box sizes ranging from $1$~Gpc to $11.2$~Gpc and mean interparticle spacings ranging from $0.27$~Mpc to $2.22$~Mpc.  Although the smaller-box simulations might be expected to underpredict power on large scales, we see that this effect is under a percent by $L=100$.  Similarly, although the simulation with the largest interparticle spacing, \FDdNSbbb{}, has significant small-scale errors, for \FDdNSbb{} we see that these only appear for $L>10000$.  The runs \FDdNS{}, \FDdNSh{}, \FDdNSb{}, and \FDdNSbb{} all agree to $1\%$ up to $L=3000$ and $<1.5\%$ beyond $L=20000$, so we may use any of them for testing \hyphi{}.  In particular, \flamingo{} simulations varying the neutrino mass and the hydrodynamic feedback parameters have the same resolution as \FDdNS{}, so Fig.~\ref{f:CLpp_convergence} confirms their suitability for our purposes.

\begin{figure}
    \begin{center}
    \includegraphics[width=90mm]{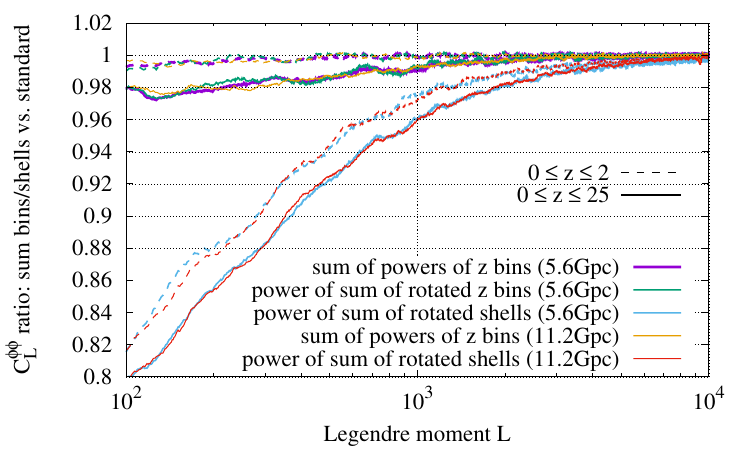}%
    \end{center}
    \caption{
    Effects of binning and bin/shell rotation on $\CLpp$.
    For \FDdNSbb{}, with a $5.6$~Gpc box, methods 
    \ref{i:sum_bins} (sum of powers of $z$ bins),
    \ref{i:sum_rot_bins} (power of sum of rotated $z$ bins), and \ref{i:sum_rot_shells} (power of sum of rotated $z$ shells) are show
    divided by the standard method \ref{i:standard}, computed over the region
    $z\leq 2$ (dashed) and $z \leq 25$ (solid).  Also shown are methods
    \ref{i:sum_bins} and \ref{i:sum_rot_shells} divided by \ref{i:standard}
    for \FDdNSbbb{}, whose large simulation volume makes it immune to 
    volume replication errors.
    \label{f:sums_rotations}
    }
\end{figure}

We consider one further issue related to \flamingo{} lightcones, that of simulation volume replication, with \FDdNSbb{} and \FDdNSbbb{} as particular examples.  The radius of a lightcone extending to $z= 25$ exceeds the simulation box size $L_{\rm sim}$ for \FDdNSbb{}, but not for \FDdNSbbb{}.  Lightcones are generated by replicating the simulation volume consistently with its periodic boundary conditions.  A photon propagating from $z=25$ to $z=0$ in \FDdNSbb{} could pass through the same structures twice, and different portions of a large-$z$ shell could see the same structures, introducing spurious correlations into our $\kappa$ map.  We consider several methods of mitigating such volume replication effects. We compute multiple $\kappa$ maps, one for each $z$ shell, and then combine them using one of four methods to obtain $\CLpp$.
\begin{enumerate}
\item Sum all $\kappa$ maps along the entire line of sight as in
    \EQ{e:kappa}, then compute $\CLpp$ from the resulting $\kappa$ map.
    This is our standard $\CLpp$ computation, used in subsequent figures
    unless noted otherwise.
    \label{i:standard}
\item Sum the $\kappa$ maps within redshift bins $z_i \leq z \leq z_{i+1}$,
    with the bin widths $\chi(z_{i+1}) - \chi(z_i)$ chosen to be of the order
    of the simulation box size.  In practice, we choose $z_0=0$, $z_1=1$, 
    $z_2=2$, $z_3=3$, $z_4=4$, $z_5=5$, $z_6=6$, $z_7=7.25$, $z_8=9.5$, 
    $z_9=12.25$, $z_{10}=15$, $z_{11}=20$, and $z_{12}=25$.  We compute a 
    separate $\kappa$ map for each bin, and a $\CLpp$ from each $\kappa$ map,
    before summing to obtain the total $\CLpp$.
    \label{i:sum_bins}
\item Compute a $\kappa$ map for each of the above $z$ bins, then rotate
    each $\kappa$ by a random angle.  The rotated $z$-binned $\kappa$ maps
    are then summed and the result used to compute $\CLpp$.
    \label{i:sum_rot_bins}
\item Compute a $\kappa$ map for each redshift shell, with $\Delta z = 0.05$
    up to $z=3$, and then randomly rotate each one. The rotated $\kappa$ maps
    are summed and the result used to compute $\CLpp$.  
    \label{i:sum_rot_shells}
\end{enumerate}
Using $\rho$ to represent a random rotation, $B$ a redshift binning, $\Sigma$ a summation over bins or shells, and $P$ the computation of an auto-power spectrum, we may represent these as: \ref{i:standard} $P(\Sigma\kappa)$; \ref{i:sum_bins} $\Sigma P(B(\kappa))$; \ref{i:sum_rot_bins} $P(\Sigma(\rho(B(\kappa))))$; and \ref{i:sum_rot_shells} $P(\Sigma \,\, \rho(\kappa))$.  Further, since the lightcone radius up to $z=2$ is less than the simulation box size for \FDdNSbb{}, we also compute $\CLpp$ integrating only up to $z=2$.  This calculation, along with both $z\leq 2$ and $z\leq 25$ light cones for \FDdNSbbb{}, are thus immune to volume replication effects.

Figure~\ref{f:sums_rotations} compares the second, third, and fourth methods to the first.  Immediately apparent is the fact that the two box sizes agree to $<1\%$ at all $L$, even though the \FDdNSbbb{} simulation does not suffer from volume replication errors.  Thus we see that these errors are negligible for our purposes, at least within the context of our power spectrum ratios smoothed with centered $100$-point moving averages.  Also apparent is the close agreement between methods \ref{i:sum_bins} and \ref{i:sum_rot_bins}, showing that bin rotation and separate-bin power spectrum computation decorrelate different $z$ bins in very similar ways.

Next, consider small scales, $L \geq 1000$, which are most relevant to tests of the non-linear $\CLpp$.  The sum-of-bins calculation, method~\ref{i:sum_bins}, and the sum-of-rotated-bins calculation, method~\ref{i:sum_rot_bins}, agree with the standard method~\ref{i:standard} to $1\%$ across this entire range, for both simulations and both light cone sizes.  Thus any of these three methods is sufficiently accurate for percent-level tests of the non-linear $\CLpp$.  Meanwhile, method~\ref{i:sum_rot_shells} exhibits significant power loss, and should not be used for testing \hyphi{}.

Finally, consider $L < 1000$ in Fig.~\ref{f:sums_rotations}.  Evidently, breaking the redshift range into a greater number of intervals, and rotating the resulting $\kappa$ maps, leads to a greater power suppression; method~\ref{i:sum_rot_shells} uses $40$ redshift shells, while methods~\ref{i:sum_bins} and \ref{i:sum_rot_bins} use $12$ redshift bins each.  Furthermore, the presence of this suppression even in the \FDdNSbbb{} power spectra and the $z\leq 2$ \FDdNSbb{} power spectrum, for which the light cone radii are less than $L_{\rm sim}$, shows that it is not due to volume replication effects.

The oscillatory nature of this low-$L$ power suppression for method~\ref{i:sum_rot_shells}, as well as its $L$-dependence, suggests that it is due to the exclusion of baryon acoustic oscillations (BAO) along the line of sight.  Since $\chi_\star/2$ is the peak of the lensing kernel of \EQ{e:g_kernel}, the Legendre moment $L$ is approximately associated with the length scale $\chi_\star/(2L) \sim 5000 L^{-1}~{\rm Mpc}/h$, which for $L\sim 50$ corresponds to the BAO scale.  Thus, not only is map rotation unnecessary for suppressing volume replication effects, but excessive rotation erroneously throws out actual large-scale correlations which ought to be included in the power spectrum.

We conclude that $\kappa$ maps from \flamingo{} lightcones, computed using method~\ref{i:standard} above for the total power and methods~\ref{i:sum_bins} or \ref{i:sum_rot_bins} for tomographic redshift bins, are suitable for testing calculations of $\CLpp$ in the non-linear regime $L\gtrsim 1000$ at the percent level.  At larger scales, $L \sim 100$, the redshift-binned methods \ref{i:sum_bins} and \ref{i:sum_rot_bins} underpredict power by $2\%$-$3\%$, an error of which we should be aware when testing the application of \hyphi{} to tomographic bins.

\subsection{Baryonic feedback}
\label{subsec:bkg:baryonic_feedback}

Stellar and AGN feedback in the \flamingo{} simulations was calibrated using machine learning to two observables: the galaxy stellar mass function (SMF) at $z=0$, measured in \citet{Driver:2022vyh}; and cluster gas fractions, $f_{\rm gas}$, measured using X-ray and weak lensing observations, as compiled in \citet{Kugel:2023wte}.  The SMF constrains the galaxy-halo connection, while $f_{\rm gas}$ correlates with the magnitude of the clustering suppression due to baryonic feedback \citep{vanDaalen:2019pst}.  This approach allowed \flamingo{} to explore the parameter space of feedback models in an observationally-relevant manner.  For example, the high-gas-fraction model \FHwdNS{} in Table~\ref{t:flamingo_sims} increased $f_{\rm gas}$ by twice the observational uncertainty, while the low-gas-fraction model \FHsdNS{} decreased $f_{\rm gas}$ by the same amount.  For a complete description of feedback in \flamingo{}, see \citet{Schaye:2023jqv,Kugel:2023wte}.

\begin{figure}
  \includegraphics[width=90mm]{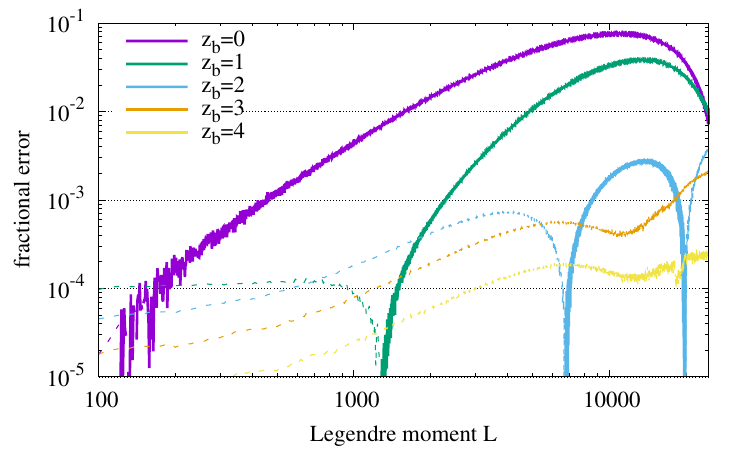}%
  \caption{
    Impact of neglecting hydrodynamic suppression of the CMB lensing potential
    power spectrum for redshifts $z\geq z_{\rm b}$.  In the $L\leq 25000$ range,
    $z_{\rm b}=1$ and $2$ lead to errors of $4\%$ and $0.4\%$, respectively.
    \label{f:impact_of_neglecting_hydro}
  }
\end{figure}

Figure~\ref{f:impact_of_neglecting_hydro} demonstrates the impact of neglecting baryonic feedback above a redshift $z_{\rm b}$ on the lensing potential power spectrum.  The figure compares \flamingo{} runs \FDdNSb{} and \FHdNSb{} from Table~\ref{t:flamingo_sims}, which are, respectively, dark-matter-only and hydrodynamic runs in $2.8$~Gpc boxes.  Neglecting feedback at all redshifts ($z_{\rm b}=0$) leads to a maximum error of $8\%$, which is reduced to $4\%$ ($0.4\%$) by including feedback for $z_{\rm b} \leq 1$ ($z_{\rm b} \leq 2$).  Thus, percent-level accuracy in the computation of $\CLpp$ up to $L \gtrsim 1000$ requires the inclusion of baryonic feedback at least at low redshifts, $z \lesssim 2$.

Effects of baryons upon the clustering of matter have been investigated using hydrodynamic simulations by \citet{vanDaalen:2011xb,vanDaalen:2013ita,vanDaalen:2019pst,Hellwing:2016ucy,Mccarthy:2017yqf,Springel:2017tpz,Chisari:2018prw,Pakmor:2022yyn}, while their impact upon CMB lensing was considered by~\citet{Chung:2019bsk,McCarthy:2021lfp}.  Recently, \citet{Salcido:2023etz} (see also \citealt{Semboloni:2011fe,vanDaalen:2019pst,Debackere:2021ado}) demonstrated that, for the purpose of determining hydrodynamic suppression of the matter power spectrum, these phenomena may be reduced to their effect on the characteristic relative baryon fractions $\fbtilhat(z) = f_\mathrm{b}(\hat M,z) / (\Ob/\Om)$ of halos of characteristic mass $\hat M(z)$ at redshift $z$.  Given $\fbtilhat$, whether measured from observations or computed through simulations, the hydrodynamic suppression factor $\Pmhyd / \Pmdmo$ is fit to $2\%$ accuracy in the range $k\leq 12~h/$Mpc and $z \leq 3$ by the \SPk{} function of that reference.

Given the full functional form of $\fbtil(M,z)$, we could in principle compute the baryonic suppression directly using a halo model.  However, this function is difficult to measure.  \citet{Salcido:2023etz} showed that, for each $z$, the $k$-dependent baryonic suppression is most strongly correlated with the baryon fraction at a single characteristic mass $\hat M(z)$.  This correlation was shown to be independent of the strength of the subgrid feedback.

Our calculations in Sec.~\ref{sec:hydrodynamic_suppression} will use $\fbtil(M,z)$ measured directly from the \flamingo{} \FHdNS{} simulations, evaluated at the $\hat M(z)$ of \citet{Salcido:2023etz}.  Qualitatively, $\fbtil$ is large for halos at high redshifts, before structures such as AGN have formed; for halos too small to form such structures; and for the largest halos, which efficiently capture baryons, and are representative of the universe as a whole. The \SPk{} fit covers the range $k \leq 12~h/$Mpc and $z\leq 3$ necessary for $\approx 2\%$-level accuracy in $\CLpp$.  Above $z=3$, we assume that the power spectrum suppression due to baryons is negligible, an approximation consistent with Fig.~\ref{f:impact_of_neglecting_hydro}.   For $k>12~h/$Mpc, we again assume a negligible suppression, an approximation which we confirm to have a sub-percent-level impact on $\CLpp$ up to $L=9000$.

\section{Non-linear CMB lensing}
\label{sec:non-linear_cmb_lensing}

\subsection{Non-linear perturbation theory}
\label{subsec:nll:non-linear_perturbation_theory}

In order to calculate the non-linear perturbative CDM+baryon (CB) power spectrum in the presence of massive neutrinos, we employ the Time-Renormalization Group (Time-RG) perturbation theory of \citet{Pietroni:2008jx,Lesgourgues:2009am}.  Time-RG integrates the non-linear continuity and Euler equations of fluid dynamics over time for each wave number, making it well-suited to massive neutrinos, which introduce a scale-dependence into the growth factor, and to dark energy models with evolving equations of state.

Let $\eta = \ln(a/\ain)$ for some initial scale factor $\ain$, and let primes denote derivatives with respect to $\eta$.  Let the perturbation indices $0$ and $1$ correspond, respectively, to the density contrast $\delta  = (\rho_{\rm CB}-\bar\rho_{\rm CB})/\bar\rho_{\rm CB}$ and the velocity divergence $\theta = -\vec\nabla\cdot\vec v / \Hc$; thus, for example, $P_{01}(k)$ represents $P_{\delta\theta}(k)$.  Then at each $k$, the Time-RG equations of motion for the CB power spectra $P_{ab}$ are
\begin{eqnarray}
  P_{ab}'
  &=&
  -\Xi_{ac} P_{cb} - \Xi_{bc} P_{ac} + I_{acd,bcd} + I_{bcd,acd},
  \label{e:trg:dP}
\end{eqnarray}
where
\begin{eqnarray}
  \left[\Xi_{ab}\right]
  &=&
  \left[
    \begin{array}{cc}
      0\quad & -1 \\
      \frac{k^2 \Phi}{\Hc^2 \delta}\quad & 1+\frac{\Hc'}{\Hc}
    \end{array}
    \right]
  \label{e:trg:Xi}
  \\
  I_{acd,bef}
  &=&
  \int_q \gamma_{acd}^{\vec k, \vec q, \vec p} B_{bef}^{\vec k, \vec q, \vec p}
  \\
  \gamma_{001}^{\vec k, \vec q, \vec p}
  &=&
  \gamma_{010}^{\vec k, \vec p, \vec q}
  =
  \frac{(\vec q + \vec p)\cdot \vec p}{2p^2}
  ~\textrm{and}~
  \gamma_{acd}^{\vec k, \vec q, \vec p}
  =
  \frac{(\vec q + \vec p)^2 \vec q \cdot \vec p}{2q^2 p^2}
  \\
  I_{acd,bef}'
  &=&
  -\Xi_{bg}I_{acd,gef} - \Xi_{eg}I_{acd,bgf} - \Xi_{fg}I_{acd,beg}+ 2A_{acd,bef}
  \label{e:trg:dI}
  \\
  A_{acd,bef}
  &=&
  \!\!\!\!\!\! \int_q \!\!\!  \gamma_{acd}^{\vec k, \vec q, \vec p}
  \Big[
    \gamma_{bgh}^{\vec k, \vec q, \vec p} P_{ge}^q P_{hf}^p
    \! + \! \gamma_{egh}^{\vec q, -\vec p, \vec k} P_{gf}^p P_{hb}^k
     \!+ \! \gamma_{fgh}^{\vec p, \vec k, -\vec q} P_{gb}^k P_{he}^q
    \Big]~~\quad
  \label{e:trg:A}
\end{eqnarray}
with all other $\gamma_{abc}$ zero, and summation over repeated indices implicit.  Here, wave number superscripts denote functional dependence, so that $P_{ab}^{\vec k}$ denotes the power spectrum $P_{ab}(\vec k)$ and $B_{abc}^{\vec k, \vec q, \vec p}$ the bispectrum $B_{abc}(\vec k, \vec q, \vec p)$.  We use $\int_q X^{\vec k, \vec q, \vec p}$ as shorthand for $\int \frac{d^3q}{(2\pi)^3}\frac{d^3p}{(2\pi)^3} (2\pi)^3 \delD(\vec k - \vec q - \vec p)  X(\vec k, \vec q, \vec p)$ for any function $X(\vec k, \vec q, \vec p)$, where $\delD$ is the three-dimensional Dirac delta function.  In Time-RG, the bipsectrum integrals $I_{acd,bef}$, initialized to zero, are the repositories of non-linear information, and are sourced by the mode-coupling integral $A_{acd,bef}$.

We may add to this system an evolution equation for the lensing potential power spectrum of \EQ{e:CLpp_los}.  Substituting $k=L/\chi$, we recast the line-of-sight integral as an integral over our time coordinate $\eta$ of
\begin{equation}
  \frac{d\CLpp}{d\eta}
  =
  \frac{9\Omo^2\Hco^4}{4 a^2 L^4 \Hc}
  \frac{g(\chi)^2}{\chi^2} P_{\rm m}\!\left(\frac{L}{\chi},z\right).
  \label{e:dCLpp_deta}
\end{equation}
where $a$, $z$, $\Hc$, and $\chi$ are now functions of $\eta$.  Since $\Pcb$, hence $\Pm$, is available at each time step in our Time-RG integration, $\CLpp$ may simply be added to our system of \EQS{e:trg:dP}{e:trg:A} with very little additional computational cost.  Furthermore, if $\Pm$ contains a hydrodynamic suppression factor parameterized by some variables, we can compute a separate $\CLpp$ for each of several choices of variables all at once, again with little additional computation.

\subsection{Non-linear lensing power}
\label{subsec:nll:non-linear_lensing power}

Next, we use non-linear perturbation theory to consider the dependence of $\CLpp$ upon clustering in different redshift ranges.  The contribution of $z_{\rm min} \leq z \leq z_{\rm max}$ is found by integrating \EQ{e:dCLpp_deta} from $\eta(z_{\rm max})$ to $\eta(z_{\rm min})$, where $\eta(z) = -\ln[(1+z)\ain]$.  We use $P_{\rm m} = (\fcb P_{00}^{1/2} + \fnu P_{\nu}^{1/2})^2$, as in \citet{Saito:2008bp}, where $P_{00}$ is the Time-RG power spectrum of \EQ{e:trg:dP}.   We shall see in the remainder of this section that Time-RG is sufficiently accurate to provide a qualitative picture of the contributions of different redshifts.

\begin{figure}
  \includegraphics[width=87mm]
                  {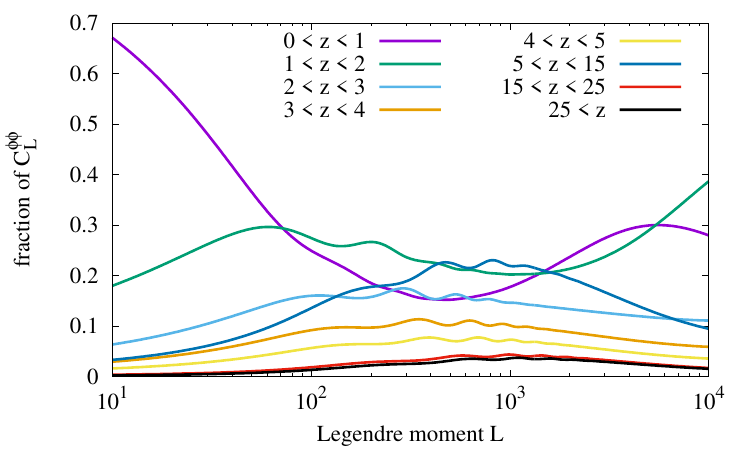}%
  \caption{
    Fractional contributions to the total lensing potential power spectrum
    $\CLpp$ due to clustering in different redshift bins.  Low $L$s
    are dominated by low-$z$ clustering, while $L\sim 1000$ receives
    significant contributions from a range of redshifts.
    \label{f:CLpp_fraction}
  }
\end{figure}

\begin{figure}
  \includegraphics[width=87mm]{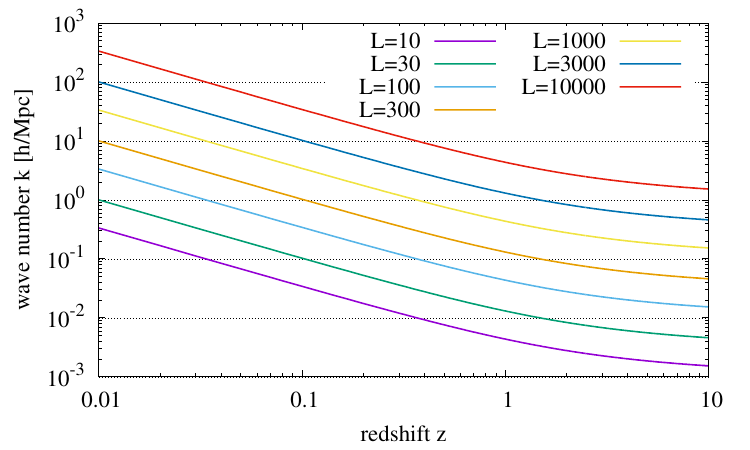}%
    \caption{
      Mapping between $L$ and $k = L/\chi(z)$ for a range of $z$.
      \label{f:ell_vs_k_z}
    }
\end{figure}

Figure~\ref{f:CLpp_fraction} shows the contributions of several redshift bins to the total lensing potential power spectrum.  The domination of the low-$L$ lensing power by low-$z$ clustering is explained by Fig.~\ref{f:ell_vs_k_z}.  Matter power spectra $P_{\rm m}(k,z)$ typically peak around $k=0.01~h/$Mpc, a scale which contributes to $\CLpp$ for $L=10$ at $z\approx 0.5$, and for $L=30$ at $z\approx 2$.  Interestingly, at $L\sim 1000$, the total contribution from $z>5$ in Fig.~\ref{f:CLpp_fraction} is $\approx 30\%$, since the lower-$z$ clustering contributing to this $L$ is drawn from increasingly large $k$, for which the power spectrum rapidly declines.

\begin{figure}
  \includegraphics[width=87mm]{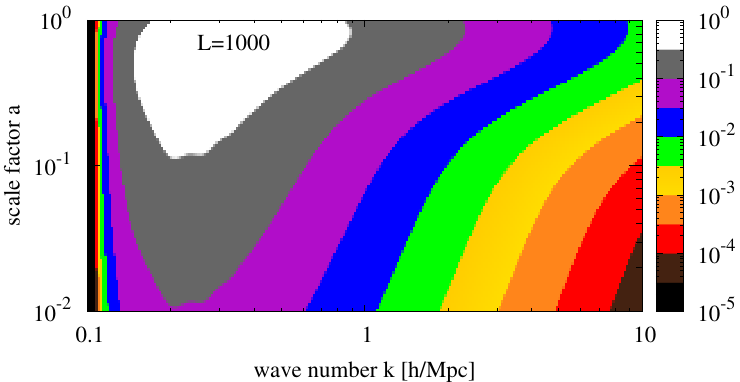}%
  
  \includegraphics[width=87mm]{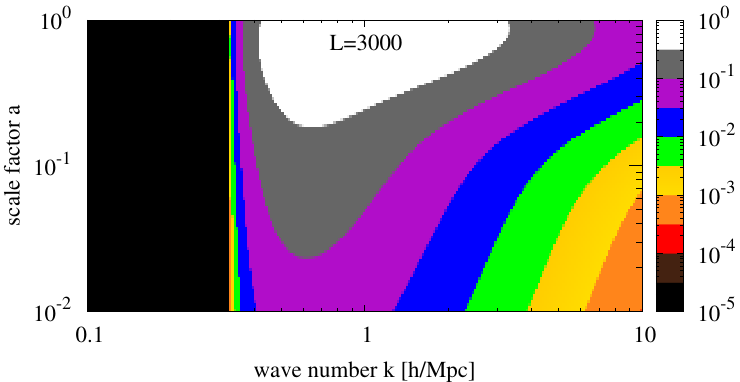}%
  \caption{
    Heat map showing $d\CLpp / d\eta / \CLpp$,
    that is, the fractional contribution of each wave number and
    scale factor, to the total $a=1$ lensing potential power spectrum.
    (Top) $L=1000$. (Bottom) $L=3000$.
    Large scales $k < L / \chi_*$ do not contribute to $\CLpp$.
    \label{f:CLpp_heat_map}
  }
\end{figure}

Heat maps quantifying the fractional contributions of different wave numbers and scale factors to $\CLpp$ are shown in Fig.~\ref{f:CLpp_heat_map}, which divides $d\CLpp/d\eta$ from \EQ{e:dCLpp_deta} by the final $a=1$ lensing potential power.  For $L=1000$ (top) and $L=3000$ (bottom), the contributions peak at $(a,k)$ of ($0.60,0.26~h/{\rm Mpc})$ and ($0.79,0.84~h/{\rm Mpc})$, respectively, and are significant within factors of $\sim 3$ of these values.  Thus, the accurate computation of $\CLpp$ up to $L$ of several thousand requires reliable estimates of the matter power spectrum for $z \lesssim 5$ and $k \lesssim 2~h/$Mpc.

\begin{figure}
  \includegraphics[width=90mm]{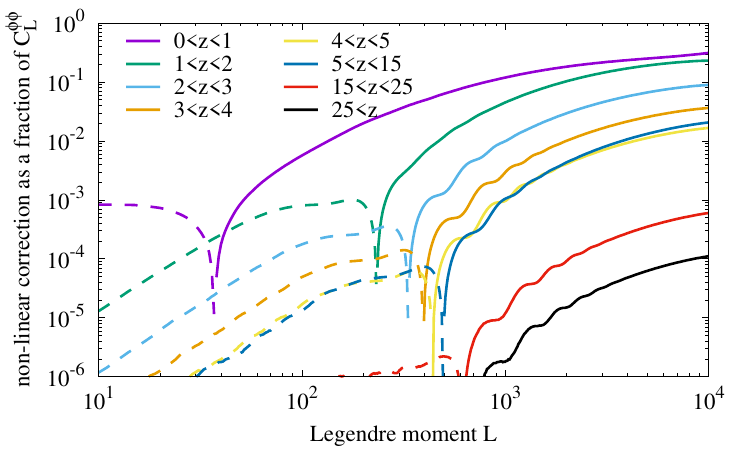}%
  \caption{
    Fractional contributions of non-linear power to the total lensing potential
    power spectrum $\CLpp$ due to clustering in different redshift bins.
    Solid (dashed) lines correspond to positive (negative) contributions.
    \label{f:CLpp_fraction_NL}
  }
\end{figure}

Since our focus is non-linear lensing at high $L$, we must also quantify the contribution of non-linear clustering to $\CLpp$, which we may do by substituting the difference between non-linear and linear matter power spectra for $P_{\rm m}$ in \EQ{e:dCLpp_deta}.  Figure~\ref{f:CLpp_fraction_NL} shows the result.  Across the entire range $L \leq 10000$ considered, the total non-linear contribution for all $z>3$ is $\leq 7\%$ of $\CLpp$, while that for all $z>5$ is $\leq 2\%$.  Thus our numerical tests of the non-linear convergence power spectrum in the remainder of this section should focus on $z \lesssim 3-5$.

\subsection{N-body fits and emulation}
\label{subsec:nll:n-body_fits_and_emulation}

Though Sec.~\ref{subsec:nll:non-linear_perturbation_theory} used the Time-RG non-linear perturbation theory to compute the matter power spectrum $P_{\rm m}(k)$, we may substitute any non-linear  $P_{\rm m}(k)$ into \EQ{e:dCLpp_deta}.  Perturbation theory accurately approximates $P_{\rm m}(k)$ at high redshifts but becomes increasingly inaccurate for $z<1$, while approximations calibrated to N-body simulations have prioritized low-redshift accuracy.

\begin{figure}
  \includegraphics[width=85mm]{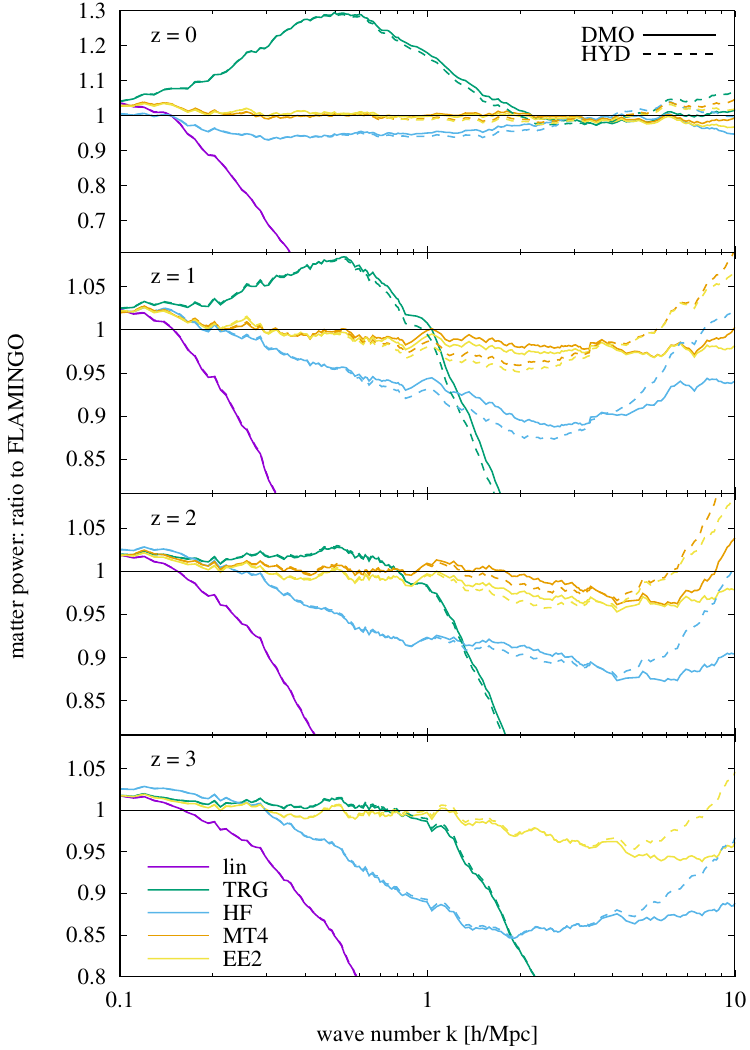}%
  \caption{
    Matter power spectrum ratios to \flamingo{} for the 
    models \FDdNSh{} (solid) and \FHdNSh{} (dashed).
    Shown are power spectra from linear (lin) and Time-RG (TRG) perturbation
    theories; the \halofit{} (HF) model of~\protect\citet{Bird:2011rb}; the
    Mira-Titan IV emulator (MT4) of~\protect\citet{Moran:2022iwe}; 
    and Euclid Emulator 2 (EE2) of ~\protect\citet{Euclid:2020rfv}.  
    \SPk was used to model the baryonic suppression.
    MT4 is limited to $z\leq 2$.
    \label{f:Pm_fid_dmo}
    \label{f:Pm_fid_hyd}
  }
\end{figure}

This work considers three such simulation-based approximations.  \halofit{}, inspired by the halo model, splits the non-linear power spectrum into a quasi-linear term, which approaches the linear power spectrum on large scales, and a non-linear term, which is fit to simulations; see~\citet{Smith:2002dz,Bird:2011rb,Takahashi:2012em}.  We implement the \halofit{} model of \citet{Bird:2011rb} which has been fit for massive neutrino cosmologies.

The second approximation is the Mira-Titan IV (MT4) emulator of \citet{Moran:2022iwe} which uses Gaussian process modelling to interpolate a suite of $111$ simulations.  Chosen carefully to span a large parameter space, this suite includes $101$ massive neutrino cosmologies with $0.00017 \leq \onu \leq 0.01$.  The third approximation, Euclid Emulator 2 (EE2) of \citet{Euclid:2020rfv}, covers a greater range of redshifts and wave numbers than the MT4 emulator, at the cost of a smaller neutrino mass range, $\sum m_\nu \leq 0.15$~eV.  \halofit{} has been calibrated up to $z=3$, the MT4 emulator to $z=2$, and EE2 to $z=10$; at higher redshifts, we revert to Time-RG in \EQ{e:dCLpp_deta}.

MT4 and EE2 are based upon simulations with mass resolutions of $10^{10}~M_{\odot}$ and $10^9~M_{\odot}$, respectively, compared with $7\times 10^9~M_{\odot}$ for standard \flamingo{} simulations, and $8\times 10^8~M_{\odot}$ for \FDdNSh{}.  MT4 reaches $k \approx 7~h/$Mpc and EE2 reaches $\approx 9~h/$Mpc, compared with $17~h/$Mpc for standard \flamingo{} simulations and $33~h/$Mpc for \FDdNSh{}.  Thus the \flamingo{} simulation suite is  appropriate for testing emulator results.  

Figure~\ref{f:Pm_fid_dmo} compares these non-linear methods at a range of redshifts.  Evidently, the two emulators provide the most accurate power spectra at all redshifts for which they are available, with the MT4 emulator slightly more accurate at $z \geq 1$ and $1~h/$Mpc~$\lesssim k \lesssim 7~h/$Mpc. As the MT4 emulator is restricted to the range $k \leq 5/$Mpc~$\approx 7~h/$Mpc, we logarithmically extrapolate its power spectrum beyond that wave number, an approximation whose accuracy evidently diminishes around $z=2$.  For $z>2$, Time-RG is somewhat more accurate than \halofit{} at large scales, $k \lesssim 2~h/$Mpc, and less accurate at small scales.

\subsection{Tests of ${\mathbf \CLpp}$}
\label{subsec:nll:tests_of_CLpp}

\begin{figure}
  \includegraphics[width=89mm]{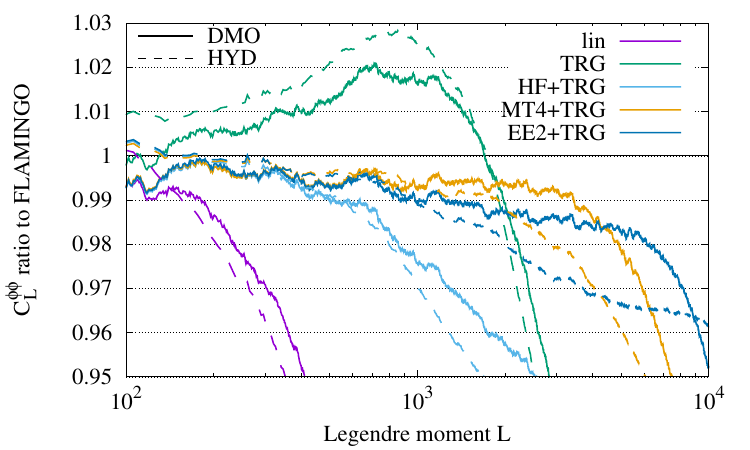}%

  \caption{
    Lensing potential power spectrum computed using linear (lin) or Time-RG
    (TRG) perturbation theories, \halofit{} (HF), the Mira-Titan IV (MT4)
    emulator,  or Euclid Emulator 2 (EE2), compared with the \flamingo{} power
    spectrum. HF is used up to $z=3$, MT4 to $z=2$, and EE2 to $z=10$, above
    which $\CLpp$ is computed using Time-RG.
    Solid lines compare DMO calculations to the \flamingo{} \FDdNSbb{} $\CLpp$
    computed up to $z=25$. 
    Dashed lines, using \SPk{} for the baryonic suppression, are compared with
    the \flamingo{} \FHdNSb{} $\CLpp$ computed to $z=5$.
    \label{f:CLpp_fid_vs_flamingo}
  }
\end{figure}

Figure~\ref{f:CLpp_fid_vs_flamingo} compares lensing potential power spectra from linear and Time-RG perturbation theories, \halofit{}, and the two emulators to the \flamingo{} lensing potential power spectrum, computed as described in Sec.~\ref{subsec:bkg:flamingo_simulations}.  As might be expected from the comparison to a higher-resolution simulation in Fig.~\ref{f:Pm_fid_dmo}, both perturbation theories underpredict small-scale power, with linear and Time-RG perturbation theory underpredicting by $\geq 3\%$ above $L\approx 300$ and $L\approx 2400$, respectively.  \halofit{} is more accurate than Time-RG for $200 \lesssim L \lesssim 1000$ but underpredicts power by $\geq 3\%$ above $L \approx 1300$.  In this and subsequent plots, the ratios to N-body power spectra are summed over all available lightcones and smoothed using centered $100$-point moving averages.

Meanwhile, the emulator-based calculations are highly accurate.  The two emulators, MT4 and EE2, agree at the percent level at $L \lesssim 6000$, with MT4 somewhat more accurate for $1000 \lesssim L \lesssim 4000$ and less accurate for $L \gtrsim 5000$.  Due to its greater neutrino mass range, and its slight accuracy advantage in Fig.~\ref{f:CLpp_fid_vs_flamingo}, we will focus henceforth on MT4.  Our standard $\CLpp$ calculation (MT4+TRG) will use the MT4 emulator for the matter power spectrum in the range $z \leq 2$ and Time-RG perturbation theory above that redshift.

\begin{figure}
  \includegraphics[width=90mm]{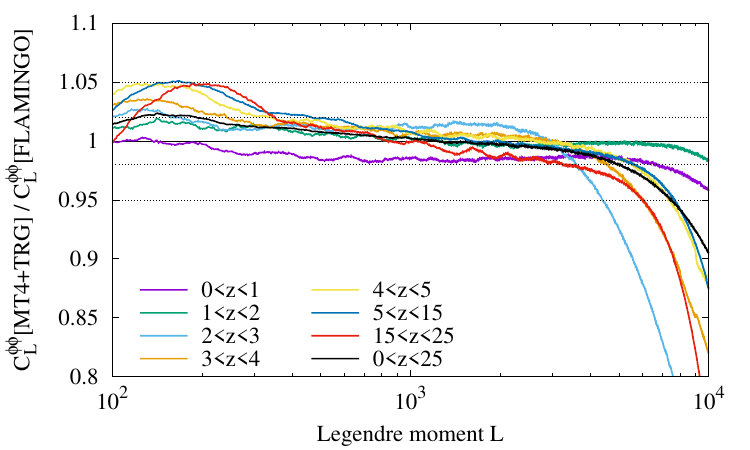}%
  \caption{
    Accuracy of $\CLpp$ contributions from several redshift bins.
    Our computation integrates \EQ{e:dCLpp_deta} over the appropriate
    redshift range, using the Mira-Titan IV Emulator for $P_{\rm m}$ at
    $z\leq 2$ and Time-RG perturbation theory at $z>2$.  This is compared
    with the \flamingo{} N-body $\CLpp$ contributions from the same
    redshift bins.  Inner and outer dotted lines show errors of $2\%$ and
    $5\%$, respectively.
    \label{f:comp_dmo_zbins}
  }
\end{figure}

The accuracy of this standard MT4+TRG calculation for matter sources in several redshift bins is assessed in Fig.~\ref{f:comp_dmo_zbins}.  Aside from low-$L$ bumps consistent with Fig.~\ref{f:sums_rotations}, our MT4+TRG calculation is accurate to $\leq 2\%$ up to $L=3000$.  For all but one redshift bin, MT4+TRG is $\leq 5\%$ accurate for $L\leq 6000$.  Thus our calculation is accurate not just for the total CMB lensing power, but also for individual redshift bins, as needed for tomographic analyses such as \citet{Peacock:2018xlz,Krolewski:2021yqy,DES:2022xxr,DES:2022urg,Wang:2022uel}.

\subsection{Massive neutrino suppression}
\label{subsec:nll:massive_neutrino_suppression}

\begin{figure}
  \includegraphics[width=87mm]{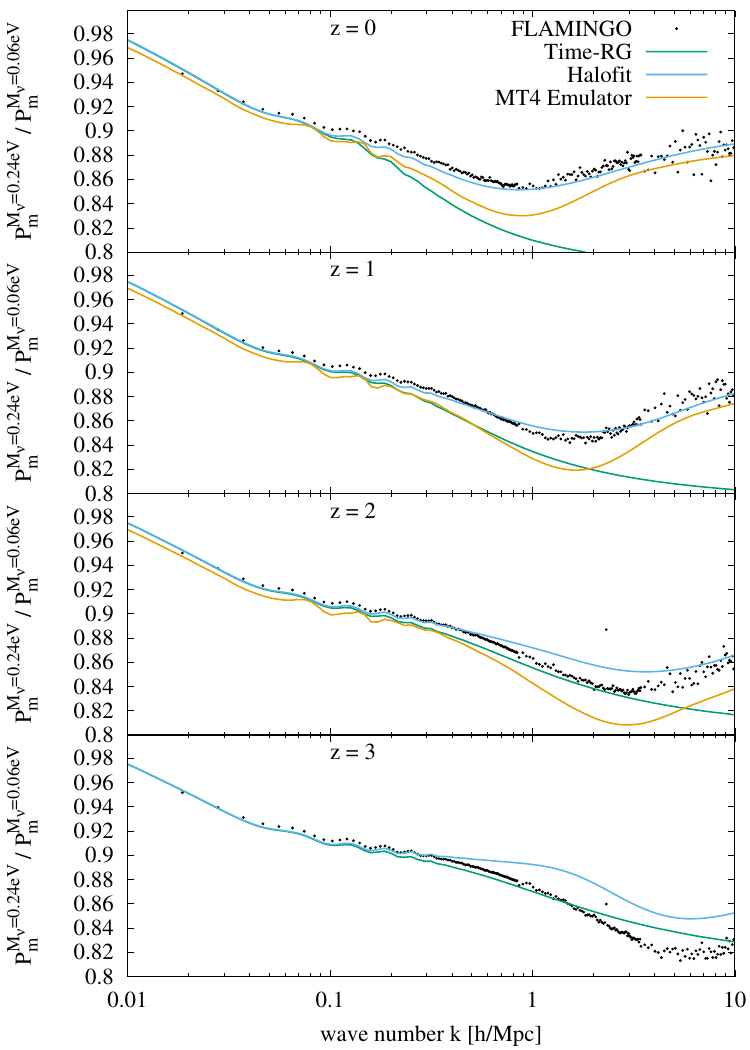}%
  \caption{
    Spoon-like feature in the ratios of power
    spectra of two models, \FDpNL{} and \FDpNS{} from
    Table~\ref{t:flamingo_sims}, differing only in their neutrino mass sums
    $M_\nu$.  \halofit{} has been calibrated to fit low-redshift spoons accurately,
    while perturbation theory is qualitatively incapable of reproducing this
    non-perturbative effect. MT4 is limited to $z\leq 2$.
    \label{f:Pm_spoon}
  }
\end{figure}

We begin by considering the effects of neutrinos on the matter power itself.  Following the approach of \citet{Lesgourgues:2009am,Upadhye:2013ndm,Upadhye:2017hdl}, we include neutrinos at the fully linear level, interpolating the ratio $\delta_\nu / \delta_{\rm cb}$ from \camb{}~\citep{Lewis:1999bs,Lewis:2002ah}.  Thus, we include neutrinos in the gravitational potential $\Phi(k,z)$ of \EQ{e:trg:Xi} as
\begin{equation}
  k^2 \Phi
  = -\frac{3}{2}\Hc^2
  \left( \Omega_{\rm cb}(z) 
  + \Omega_\nu(z) \left.\frac{\delta_\nu}{\delta_{\rm cb}}\right|_{\rm lin}\right)
  \delta_{\rm cb}.
  \label{e:Phi_nuLin}
\end{equation}

Ratios of matter power spectra differing only in $M_\nu$, with $\Omo$, $\Obo$, $h$, $n_s$, $A_s$, $w_0$, and $w_a$ held fixed, exhibit a characteristic ``spoon'' feature shown in Fig.~\ref{f:Pm_spoon} that even non-linear perturbation theory fails to capture.  \citet{Hannestad:2020rzl} showed this to be a consequence of halo formation.  \halofit{}, which was calibrated in \citet{Bird:2011rb} to fit this feature, agrees closely with the N-body ratio, at least at low $z$, while the Mira-Titan IV emulator is accurate to a few percent.

\begin{figure}
  \includegraphics[width=85mm]{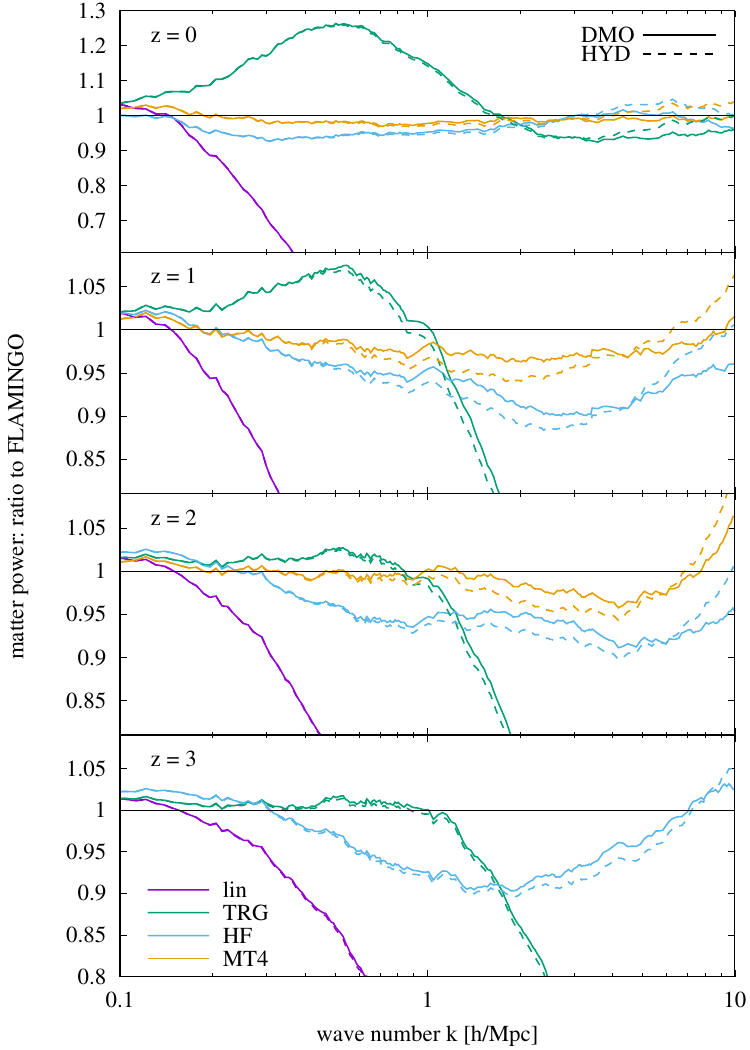}      
  \caption{
    Matter power spectrum ratios to \flamingo{} for the 
    models \FDpNL{} (solid) and \FHpNL{} (dashed), with $M_\nu = 0.24$~eV.
    Shown are power spectra from linear (lin) and Time-RG (TRG) perturbation
    theories, \halofit{} (HF), and the Mira-Titan IV emulator (MT4), 
    with \SPk used to model the baryonic suppression.
    MT4 is limited to $z\leq 2$.
    \label{f:Pm_planckLargeNuFix_dmo}
    \label{f:Pm_planckLargeNuFix_hyd}
  }
\end{figure}

Figure~\ref{f:Pm_planckLargeNuFix_dmo} plots the matter power spectrum for a massive neutrino cosmology with $M_\nu=0.24$~eV.  In spite of \halofit{}'s accurate computation of the neutrino spoon for $z<2$, we find that the matter power itself is more accurately computed for $z\leq 2$ by the Mira-Titan IV emulator, and for $z>2$ at large scales, $k \leq 1~h/$Mpc, by Time-RG perturbation theory.  Thus, our MT4+TRG calculation continues to be the best approximation to the matter power spectrum for the $k$ and $z$ relevant to computing the CMB lensing power spectrum in $\nu\Lambda$CDM cosmologies.

\begin{figure}
  \includegraphics[width=87mm]{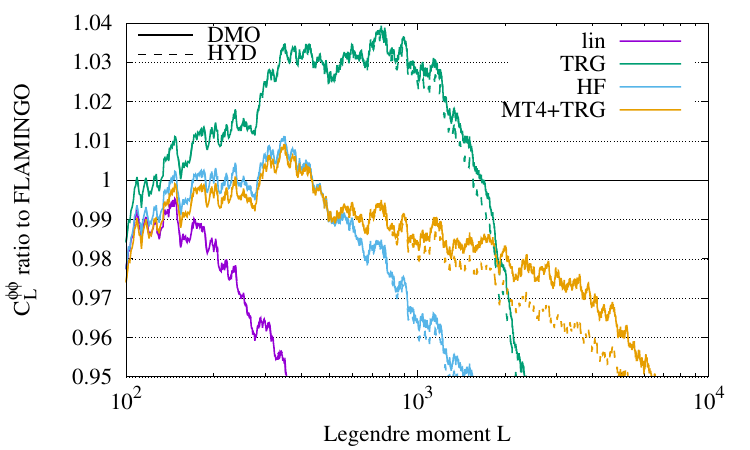}%
  \caption{
    Lensing potential power spectra for $M_\nu=0.24$~eV, computed using linear
    (lin) or Time-RG (TRG) perturbation theories, \halofit{} (HF), or the
    Mira-Titan IV (MT4) emulator, compared with \flamingo{} models \FDpNL{} and
    \FHpNL{}, computed using lightcone sources up to $z=3$.  For the
    orange curves, MT4 is used up to $z=2$ and TRG above that.  
    Solid lines compare DMO $\CLpp$ to \FDpNL{}.
    Dashed lines compare $\CLpp$ from the above methods, using 
    using \SPk{} for the baryonic suppression, to \FHpNL{}.
    \label{f:CLpp_planckLargeNuFix_dmo}
  }
\end{figure}

Next, we return to our computation of the lensing potential power spectrum in massive neutrino models.  We tested our fully-linear neutrino approximation by allowing neutrinos to respond linearly to the non-linear CB growth, using the code of \citet{Chen:2020bdf}, and found the impact on $\CLpp$ to be $<0.06\%$ even for neutrino fractions as large as $\Ono h^2 = 0.01$.  \citet{Chen:2022cgw} used a non-linear perturbation theory for massive neutrinos to show that the non-linear neutrino power is at most $2$-$3$ times the linear response power for $\Ono h^2 \leq 0.005$, corresponding to $M_\nu=0.47$~eV, so we may safely bound the impact of neutrino non-linearity to $<0.2\%$ over this range.  This is a negligible source of error for near-future experiments, so we use the fully-linear-$\nu$ approximation of \EQ{e:Phi_nuLin} henceforth.

Errors in the CMB lensing potential power spectra for a model with $M_\nu=0.24$~eV are shown in Fig.~\ref{f:CLpp_planckLargeNuFix_dmo}.  The \flamingo{} simulation uses a $1000$~Mpc box, for which we have lightcones up to $z=3$, so the other calculations are integrated only to $z=3$ for comparison.  In accordance with our previous results, the MT4+TRG calculation is accurate to $<2.5\%$ up to $L=2000$ and to $<5\%$ up to $L=6000$.  These error estimates are limited by noise in the simulation itself, since only one high-$z$ lightcone is available in the high-$M_\nu$ run.

\section{Hydrodynamic suppression}
\label{sec:hydrodynamic_suppression}

\subsection{Matter and convergence power spectra}
\label{subsec:hyd:matter_and_convergence_power_spectra}

Dashed curves in Fig.~\ref{f:Pm_fid_hyd} compare the total matter power spectrum from the high-resolution \flamingo{} hydrodynamic simulation, \FHdNSh{}, to approximations applying the \SPk{} fitting function of \citet{Salcido:2023etz} to each of the following: linear and Time-RG perturbation theories, \halofit{}, and MT4 and EE2 emulators.  Since \SPk{} applied to $P_{\rm m}$ is a multiplicative correction, the ratio of corresponding dashed and solid lines is the same in all cases, and the deviation between each dashed and solid line is due to error in \SPk{}.  At all redshifts, this error is $\leq 2\%$ up to $k\approx 5~h/$Mpc.

Aside from the small systematic overprediction of power at small scales, $k \gtrsim 5~h/$Mpc, for the hydro models, the DMO and hydro curves are qualitatively similar.  Thus, our earlier conclusions apply as well to the hydro power spectra of Fig.~\ref{f:Pm_fid_hyd}: the two emulators predict the power spectrum accurately across their tested redshift ranges, while above $z=2$, Time-RG perturbation theory accurately calculates the power up to $k \approx 1~h/$Mpc and \halofit{} above $k \approx 2~h/$Mpc.

Each of these methods, along with the \SPk{} fitting function, has been applied to the computation of the CMB lensing potential power spectrum in the dashed curves of Fig.~\ref{f:CLpp_fid_vs_flamingo}.  They are compared to the \flamingo{} hydrodynamic simulation, \FHdNSb{}, whose lightcones cover redshifts $z \leq 5$.  Our standard MT4+TRG+\SPk computation is $2\%$ accurate up to $L=3000$ and $5\%$ accurate to $L=6000$, while the EE2+\SPk computation, which reaches smaller scales, is $4\%$ accurate all the way to $L=10000$.  Figure~\ref{f:CLpp_planckLargeNuFix_dmo} shows a similar accuracy level, $3\%$ to $L=2000$ and $5\%$ to $L=5000$, for a model with both baryonic feedback and massive neutrinos, \FHpNL{}.

\begin{figure}
  \includegraphics[width=87mm]{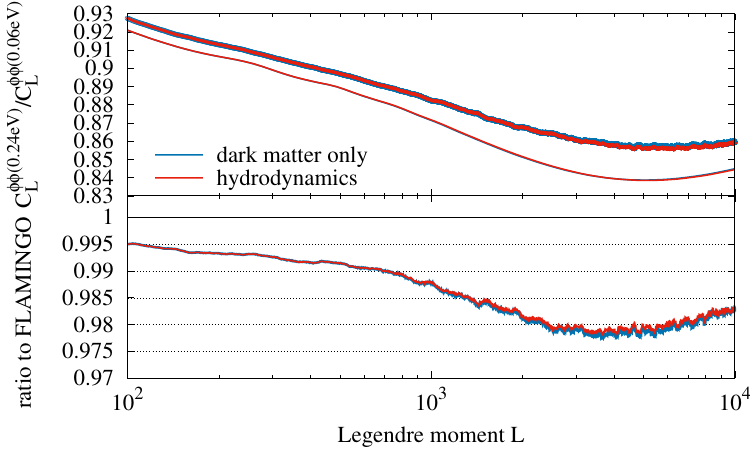}%
  \caption{
    Spoon feature in the ratio of high-$M_\nu$ to low-$M_\nu$ power spectra.
    (Top) $\CLpp$ ratio. Thick lines show the \flamingo{} (\FDpNL{}/\FDpNS{} and
    \FHpNL{}/\FHpNS{}) spoons, while thin lines show the corresponding MT4+TRG 
    spoons. 
    (Bottom) Ratio of each MT4+TRG spoon to its \flamingo{} counterpart.
    \label{f:CLpp_spoon}
  }
\end{figure}

Figure~\ref{f:CLpp_spoon} plots neutrino spoons for the lensing potential power spectra, analogous to Fig.~\ref{f:Pm_spoon} for the matter power.  The power spectrum ratio is less sensitive to hydrodynamic effects than the power spectrum itself, which is why the dark-matter-only and hydrodynamic curves are nearly identical.  Furthermore, this ratio also reduces errors present in perturbation theory and the emulator, so that the MT4+TRG spoon is correct to $<2.5\%$ for all $L\leq 10000$.

Nevertheless, we may wonder about the impact of the systematic $\approx 2\%$ overprediction of the spoon depth seen in Fig.~\ref{f:CLpp_spoon}.  While this would bias neutrino mass measurements from the spoon feature, at present the $\CLpp$ spoon is not a reliable observable.  Firstly, this is because the upturn in the spoon occurs only after $L=8000$, well beyond current observational capabilities.  Secondly and more fundamentally, the spoon itself is a comparison between the real, observed universe, whose cosmological parameters we do not know, and a hypothetical universe differing only in $M_\nu$, significantly complicating its use as an observable.  Theoretical investigations of the spoon, along the lines of \citet{Hannestad:2020rzl}, may prefer to use \halofit{} rather than MT4+TRG.

\subsection{Factorizability of suppressions}
\label{subsec:hyd:factorizability}

Since neutrinos and baryonic feedback both suppress small-scale clustering, one may worry that errors in our feedback approximations will lead to significant biases in the neutrino mass determination.  However, \citet{Mummery:2017lcn}, using the {\tt BAHAMAS} simulations of \citet{McCarthy:2016mry,Mccarthy:2017yqf}, demonstrated for several clustering statistics that neutrino and baryonic suppressions factorize.  That is, the combined effect of neutrino and baryon suppression is the product of the individual suppression factors.  Here we show that this factorization of neutrino and baryon effects also applies to the \flamingo{} CMB lensing potential power spectrum.

\begin{figure}
  \includegraphics[width=90mm]{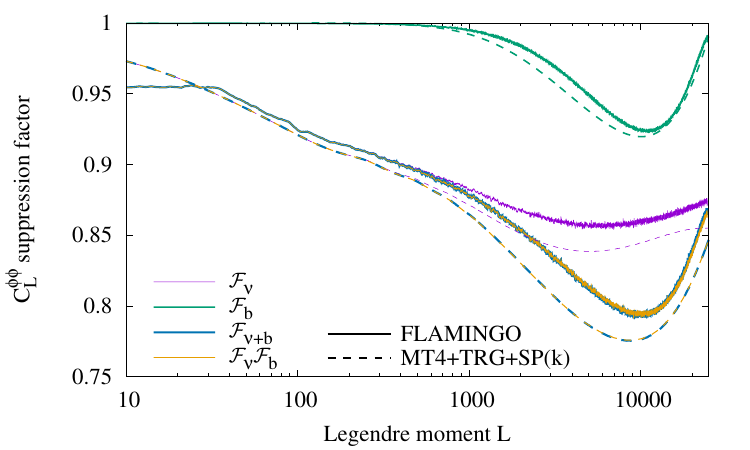}%
  \caption{
    Factorizability of neutrino and hydrodynamic suppressions of the
    CMB lensing potential power spectrum.
    \flamingo{} simulations (solid) and MT4+TRG calculations (dashed) of
    the neutrino suppression factor ${\mathcal F}_\nu$,
    the baryon suppression factor ${\mathcal F}_\mathrm{b}$,
    the combined suppression factor ${\mathcal F}_{\nu+\mathrm{b}}$,
    and the product of the first two, which accurately approximates the third.
    \label{f:factorization}
  }
\end{figure}

Consider the Planck-like cosmologies of \FDpNS{}, \FDpNL{}, \FHpNS{}, and \FHpNL{} from Table~\ref{t:flamingo_sims}, which have either a minimal ($M_\nu=0.06$~eV) or high ($M_\nu=0.24$~eV) neutrino mass sum, and are simulated using dark matter only (dmo) or a hydrodynamic (hyd) simulations.  We may define three different suppression factors, capturing the effects of neutrinos, baryons, and both at the same time:
\begin{eqnarray}
  {\mathcal F}_\nu
  &=&
  \frac{
  \CLpp({\rm dmo},M_\nu\!=\!0.24{\rm eV})
  }{\CLpp({\rm dmo},M_\nu\!=\!0.06{\rm eV})}
  \\
  {\mathcal F}_\mathrm{b}
  &=&
  \frac{
  \CLpp({\rm hyd},M_\nu\!=\!0.06{\rm eV})
  }{ \CLpp({\rm dmo},M_\nu\!=\!0.06{\rm eV})}
  \\
  {\mathcal F}_{\nu+\mathrm{b}}
  &=&
  \frac{
  \CLpp({\rm hyd},M_\nu\!=\!0.24{\rm eV})
  }{ \CLpp({\rm dmo},M_\nu\!=\!0.06{\rm eV})}
\end{eqnarray}
Factorizability then implies that ${\mathcal F}_{\nu+\mathrm{b}} = {\mathcal F}_\nu \times {\mathcal F}_\mathrm{b}$.

Figure~\ref{f:factorization} demonstrates the factorizability of neutrino and baryon suppressions in the CMB lensing potential power spectrum, using simulations as well as our standard MT4+TRG+\SPk calculation.  While the simulations and emulators differ by $\approx 2\%$ on small scales, in keeping with our previous results, in each case factorizability of the neutrino and baryon suppressions is accurate to  better than $1\%$.  

This is due to the fact that the two suppressions depend very differently on length and time.  Neutrino free-streaming suppresses clustering below the free-streaming length, corresponding to $k_\mathrm{FS} \lesssim 0.1~h/$Mpc for typical masses, which is manifested in a $\sim 10\%$ suppression of $\CLpp$ even at $L=100$.  Meanwhile, baryonic suppression is negligible below $L=1000$ and is most significant around $L=10000$.  Further, the neutrino free-streaming length is larger at earlier times, and the resulting impact upon $\CLpp$ occurs over a wide range of redshifts, as opposed to the baryonic suppression, which was shown in Fig.~\ref{f:impact_of_neglecting_hydro} to be dominated by $z \lesssim 1$.

\begin{figure}
  \includegraphics[width=90mm]{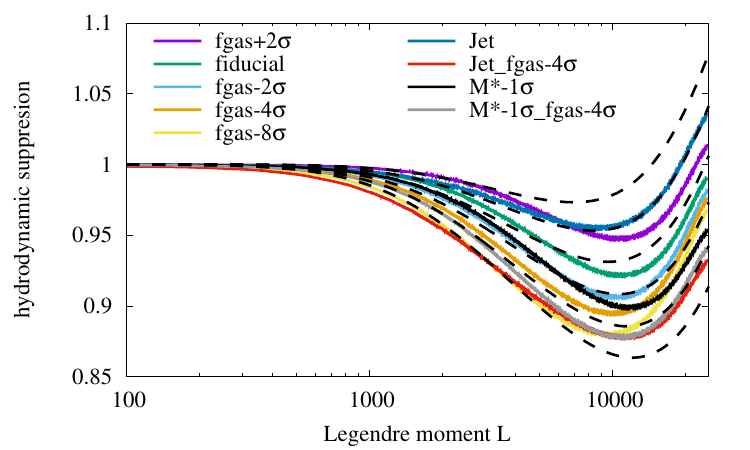}%
  \caption{
    Our $p_f$ feedback parameterization (dashed),
    which raises the \FHdNS{} baryon fraction $\fbtilhat$
    to the power $p_f$ before applying the \SPk fit of \citet{Salcido:2023etz},
    is compared to a wide range of \flamingo{} hydrodynamic suppression factors 
    (solid).  Dashed curves range from $p_f=0.5$ (uppermost curve) to $p_f=1.5$ 
    (lowest curve) in increments of $\Delta p_f = 0.2$.
    \label{f:pftb_vs_flamingo}
  }
\end{figure}

The reader may wonder how well such factorizability applies to very different hydrodynamic feedback models.  While \SPk was calibrated to a wide range of feedback models \citep{Salcido:2023etz}, a thorough exploration of the parameter space is beyond the scope of this article.  For simplicity, consider a one-parameter family of generalizations which raise $\fbtilhat(z) = f_\mathrm{b}(\hat M,z) / (\Ob/\Om)$ to a non-negative power $p_f$ before applying \SPk.  In this case, $p_f=0$ implies no hydrodynamic feedback; $p_f<1$ implies feedback that is weaker than that of the \flamingo{} fiducial model; and $p_f>1$ to feedback that is stronger.  Figure~\ref{f:pftb_vs_flamingo} compares this approximation to a wide range of \flamingo{} feedback methods and finds $0.5 \leq p_f \leq 1.5$ to cover this range, to $L\approx 4000$, aside from one model, \FHjsdNS{}.

\begin{figure}
  \includegraphics[width=90mm]{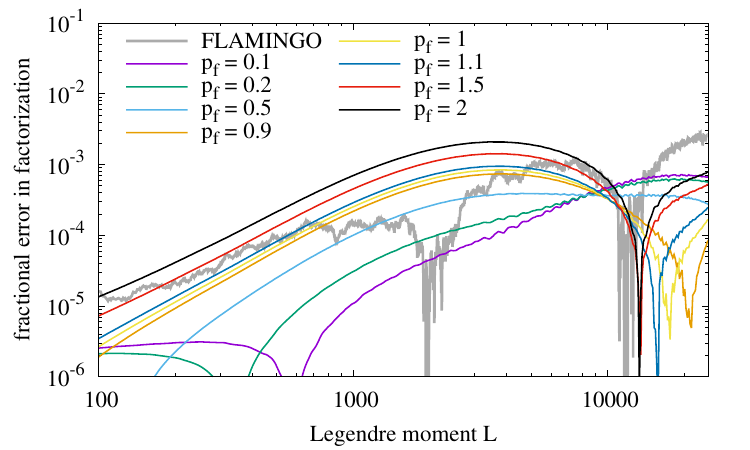}%
  \caption{
    Error in factorizability 
$|{\mathcal F}_\nu\times{\mathcal F}_\mathrm{b}/{\mathcal F}_{\nu+\mathrm{b}}-1|$
    vs. $p_f$ in the generalized feedback model
    raising $\fbtil(\hat M,z)$ to the power $p_f$ in \SPk{}. Here, $p_f<1$
    ($p_f>1$) corresponds to weaker (stronger) feedback than the \flamingo{}
    fiducial model.  Also shown is the \flamingo{} factorizability error.
    \label{f:vary_pftb}
  }
\end{figure}

Figure~\ref{f:vary_pftb} shows that factorizability of $\CLpp$ holds to excellent precision for a wide range of $p_f$.  While errors at $100 \lesssim L \lesssim 6000$ tend to rise with $p_f$, they remain $\leq 0.2\%$ all the way to $p_f=2$, which has substantially stronger feedback, and are $<0.1\%$ at all $L$ for the fiducial case, $p_f=1$.  Also shown is the factorization error in \flamingo{} itself, which is $\leq 0.3\%$ everywhere and has a similar functional form to \hyphi{} results.  Factorizability opens up the possibility of a new \SPk-like feedback fit applying directly to $\CLpp({\rm hyd}) / \CLpp({\rm dmo})$, whose parameters could be treated as nuisance parameters whose marginalization would reveal the underlying neutrino suppression.

While \citet{Salcido:2023etz} found that factorizability is a good approximation at the $2\%$ level with a history-independent baryonic correction model, the possibility remains that factorizability fails below this error threshold, or that baryonic correction is somewhat history-dependent.  This limits the bound that we can place on the accuracy of factorizability, though we note that the direct factorizability measurement from \flamingo{} in Fig.~\ref{f:vary_pftb} is consistent with our result of sub-percent-level factorization errors.

\subsection{Applicability to data constraints}
\label{subsec:hyd:applicability_to_data_constraints}

Complete forecast constraints such as those of \citet{McCarthy:2021lfp}, but using \SPk through \hyphi{}, would first require forecast constraints on the halo baryon fraction $\fbtilhat(z)$ as a function of redshift.  As such, they are beyond the scope of this article.  However, before concluding, we comment on the applicability of \hyphi{} to data and forecast constraints.

The simplest, but least powerful, approach is to parameterize $\fbtil(M,z)$ and then to treat these parameters as nuisance parameters over which to marginalize.  \citet{Salcido:2023etz} explores two- and three-parameter approximations to $\fbtil(M,z)$.  A one-parameter approach, described in Sec.~\ref{subsec:hyd:factorizability}, begins with a representative $\fbtil(M,z)$ measured from a hydrodynamic simulation, and raises it to a power $p_f$ prior to applying \SPk, with $p_f$ marginalized over as a nuisance parameter.  Figure~\ref{f:pftb_vs_flamingo} demonstrates that the range $0.5 \leq p_f \leq 1.5$ covers a wide range of feedback models.

However, marginalizing over broad priors on $\fbtil(M,z)$ offers no significant advantages over directly parameterizing $\Pmhyd(k)/\Pmdmo(k)$ as a function of subgrid feedback parameters, as done by \citet{Mead:2020vgs} in {\textsc{HMcode}}. As such, we anticipate that the resulting forecast constraints will be similar to those of \citet{McCarthy:2021lfp}.

The true strength of \SPk-based power spectrum computations such as \hyphi{} is that $\fbtil(M,z)$ is a physical observable.  Measurements of the baryon fraction may be used to narrow the range over which $\fbtil(M,z)$ are marginalized.  For example, \citet{Salcido:2023etz} parameterizes $\fbtil(M,z)$ and then provides parameter ranges consistent at the $2\sigma$ and $3\sigma$ level with the low-redshift measurements of \citet{Akino:2021ole}.  As CMB lensing is sensitive to baryonic suppression at higher redshifts, $z \sim 1$, SZ cluster constraints such as those of \citet{DES:2021sgf} also provide useful priors on the baryon fraction, with the caveat that SZ surveys are sensitive to halos with masses larger than the $\hat M(z)$ used in \SPk.

\section{Conclusion}
\label{sec:conclusion}

Weak lensing of the cosmic microwave background is becoming increasingly important as a galaxy-bias-independent constraint on the sum of neutrino masses $M_\nu$.  Our ability to interpret upcoming lensing measurements at small scales depends crucially on our ability to distinguish between the scale-dependent clustering suppressions due to baryons and neutrino free-streaming in the non-linear regime.  Employing the \flamingo{} suite of simulations, the largest-particle-number hydrodynamic simulations reaching $z=0$ to date, we systematically compared several methods for computing the non-linear CMB lensing potential power spectrum, implemented in our \hyphi{} code.

Our preferred method combined the Mira-Titan IV DMO power spectrum emulator at $z\leq 2$ with Time-RG perturbation theory at higher redshifts.  Comparing \hyphi{} with a $5.6$~Gpc-box DMO simulation in Fig.~\ref{f:CLpp_fid_vs_flamingo}, we found agreement to $1\%$ up to $L=3000$ and $2\%$ to $L=5000$.  Figure~\ref{f:comp_dmo_zbins} shows that \hyphi{} attained $5\%$ accuracy to $L=4000$ in individual redshift bins, applicable to tomographic analyses.  Including more massive neutrinos, as well as baryonic feedback through the \SPk{} fitting function, individually or simultaneously, as in Figs.~\ref{f:CLpp_fid_vs_flamingo} and \ref{f:CLpp_planckLargeNuFix_dmo}, slightly degraded the accuracy of \hyphi{}, but even with both effects, we found MT4+TRG+\SPk to be $4\%$ accurate up to $L=4000$.  Finally, we demonstrated that neutrino and baryonic suppression effects on $\CLpp$ factorize, potentially facilitating the marginalization over baryonic effects in the future.

\section*{Acknowledgements}

This project has received funding from the European Research Council (ERC) under the European Union’s Horizon 2020 research and innovation programme (grant agreement No 769130).  WE acknowledges STFC Consolidated Grant ST/X001075/1.  RK acknowledges funding by Vici grant 639.043.409, Veni grant 639.041.751 and research programme Athena 184.034.002 from the Dutch Research Council (NWO). 
CSF acknowledges support from ERC Advanced Investigator grant, DMIDAS [GA
786910].  This paper makes use of the DiRAC Data Centric system at Durham
University, operated by the Institute for Computational Cosmology on
behalf of the STFC DiRAC HPC Facility ({\tt http://www.dirac.ac.uk}). This equipment
was funded by BIS National E-infrastructure capital grant ST/K00042X/1,
STFC capital grants ST/H008519/1 and ST/K00087X/1, STFC DiRAC Operations
grant ST/K003267/1 and Durham University. DiRAC is part of the National
E-Infrastructure.

\section*{Data Availability}

The \hyphi{} code is available at {\tt github.com/upadhye/hyphi}~.  The authors of \citet{Schaye:2023jqv} intend to make the \flamingo{} simulation data public at a later date.

\appendix
\section{Code usage and performance}
\label{sec:code_usage_and_performance}

\begin{figure}
  \includegraphics[width=90mm]{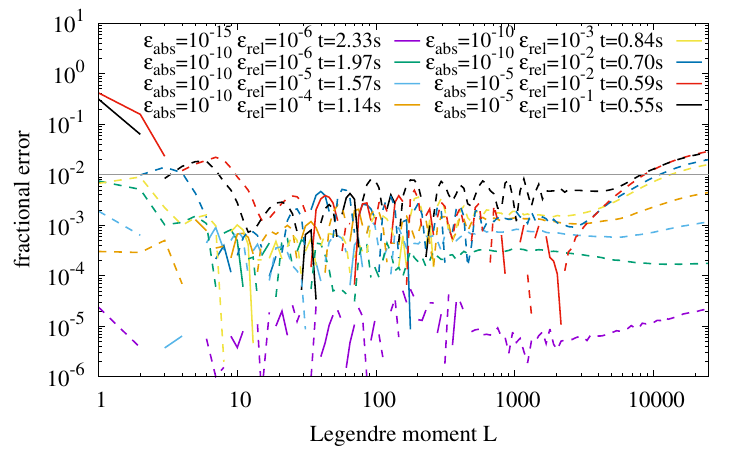}%
  \caption{
    Numerical convergence of \hyphi{} $\CLpp$ computation.  Each $\CLpp$
    computation, labeled by its absolute and relative tolerances as well as
    its running time on a standard desktop computer, is compared to a
    high-quality run with absolute and relative error tolerances
    $\epsilon_{\rm abs}=10^{-20}$ and $\epsilon_{\rm rel}=10^{-12}$, 
    respectively, whose running time is $26.16$~s.  
    \label{f:hyphi_precision}
  }
\end{figure}

\begin{figure}
  \includegraphics[width=87mm]{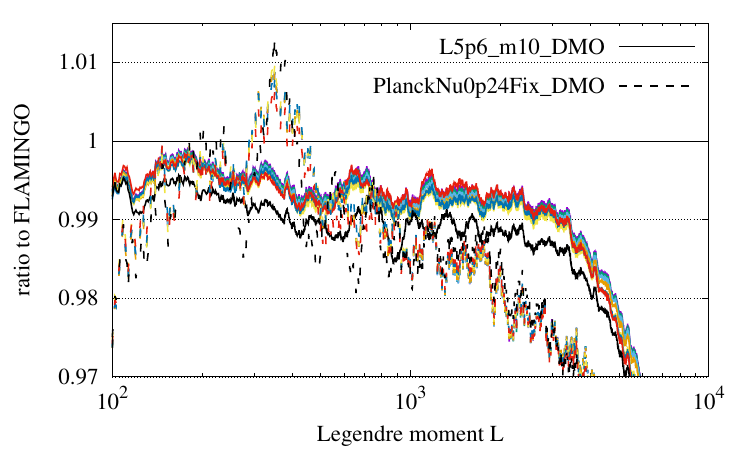}%
  \caption{
    Accuracy of \hyphi{} $\CLpp$ computation compared with the \flamingo{}
    \FDdNSbb{} (solid) and \FDpNL{} (dashed) simulations.  Absolute and relative
    tolerances for each line color are the same as in Fig.~\ref{f:hyphi_precision}.
    \label{f:hyphi_accuracy_vs_flamingo}
  }
\end{figure}

Our CMB lensing potential power spectrum code, \hyphi{}, is publicly available at {\tt github.com/upadhye/hyphi}~.  Its precision and accuracy are respectively shown in Figs.~\ref{f:hyphi_precision} and \ref{f:hyphi_accuracy_vs_flamingo}, for several combinations of the absolute and relative error tolerances $\epsilon_{\rm abs}$ and $\epsilon_{\rm rel}$, along with the running time for a single $\CLpp$ computation on a standard desktop computer. Evidently, a highly accurate lensing potential power spectrum may be computed in a fraction of a second.   

Figure~\ref{f:hyphi_accuracy_vs_flamingo} also demonstrates the speed and accuracy of \hyphi{} for larger neutrino masses.  Since only one high-redshift lightcone is available, the simulated $\CLpp$ is noisier, preventing us from quantifying an accuracy better than $2\%-3\%$.  Evidently \hyphi{} is $\leq 3\%$ accurate to nearly $L=4000$.  The running time for each $\epsilon_{\rm abs}$ and $\epsilon_{\rm rel}$ combination is within $30\%$ of the values listed in Fig.~\ref{f:hyphi_precision}.

A technical point in the \FDpNL{} computation deserves further comment.  The \flamingo{} \FDpNL{} lightcone covers the redshift range $0 \leq z \leq 3$, so we compare it with the \hyphi{} $\CLpp$ including only lens masses from that range of redshifts.  Although \hyphi{} can begin integrating $\CLpp$ at $z=3$, we find that doing so accurately requires a low absolute error tolerance, and correspondingly longer running times.  Instead, in the dashed curves of Fig.~\ref{f:hyphi_accuracy_vs_flamingo}, we have integrated $\CLpp$ over all redshifts, but printed two separate power spectra at $z$ of $3$ and $0$.  These are then differenced to obtain the power spectra shown in the figure, allowing us to achieve running times comparable to those of \FDdNSbb{}.  This technique is also applicable to tomographic analyses.  For example, printing $\CLpp$ at redshifts $3$, $2$, $1$, and $0$, then differencing successive outputs, will provide fast and accurate power spectra in the $2\leq z \leq 3$, $1 \leq z \leq 2$, and $0 \leq z \leq 1$ bins, respectively.

Each \hyphi{} run requires an input transfer function for the purpose of normalizing its matter power spectrum.  Furthermore, the linear neutrino density as a function of time is interpolated from transfer functions over a range of redshifts.  We find accurate results using transfer functions at the following $12$ redshifts: $200$, $100$, $50$, $20$, $10$, $5$, $4$, $3$, $2$, $1$, $0.5$, and $0$.  Currently, \hyphi{} is written to accept the $13$-column transfer function files output by current versions of the \camb{} code.  Additionally, in its most accurate setting, \hyphi{} uses the Mira-Titan IV emulator, which must be compiled separately.  Further instructions for compilation and usage of \hyphi{} are available at the web site listed above.

Two options are provided for implementing hydrodynamic corrections in \hyphi{} through \SPk{}.  Firstly, the baryon fraction for the BAHAMAS simulations of \citet{McCarthy:2016mry,Mccarthy:2017yqf} is included with the code, and may be raised to a positive power $p_f$ as in Sec.~\ref{subsec:hyd:factorizability} prior to application of \SPk{}.  Secondly, a variant of the power-law baryon fraction approximation of \citet{Akino:2021ole}, as implemented in \SPk{} by \citet{Salcido:2023etz}, is included in \hyphi{}.  Its three parameters determine the normalization, mass-dependence, and redshift-dependence of the baryon fraction.

\bibliographystyle{mnras}
\bibliography{hyphi}
\bsp
\label{lastpage}
\end{document}